\documentclass[11pt,a4]{article}

\usepackage{bm,amsmath,amssymb,cite,hyperref,a4wide,subfigure,graphicx}

\begin{document}

\title{
\textbf{
Spin-flavor oscillations of Dirac neutrinos in a plane electromagnetic wave}
}

%\author{Maxim Dvornikov}
%\email{maxdvo@izmiran.ru}
%
%\affiliation{Pushkov Institute of Terrestrial Magnetism, Ionosphere
%and Radiowave Propagation (IZMIRAN),
%108840 Moscow, Troitsk, Russia}
%
%\affiliation{Physics Faculty, National Research Tomsk State University, 36 Lenin Avenue, 634050 Tomsk, Russia}
%
%\affiliation{Nordita, 17 Roslagstullsbacken, 10691 Stockholm, Sweden}
%
%\date{\today}

\author{Maxim Dvornikov$^{a,b,c}$\thanks{maxdvo@izmiran.ru}
\\
$^{a}$\small{\ Pushkov Institute of Terrestrial Magnetism, Ionosphere} \\
\small{and Radiowave Propagation (IZMIRAN),} \\
\small{108840 Troitsk, Moscow, Russia;} \\
$^{b}$\small{\ Physics Faculty, National Research Tomsk State University,} \\
\small{36 Lenin Avenue, 634050 Tomsk, Russia,} \\
$^{c}$\small{\ Nordita, 17 Roslagstullsbacken, 10691 Stockholm, Sweden}}

\date{}

\maketitle

\begin{abstract}
We study spin and spin-flavor oscillations of Dirac neutrinos in a
plane electromagnetic wave with circular polarization. The evolution
of massive neutrinos with nonzero magnetic moments in the field of
an electromagnetic wave is based on the exact solution of the Dirac-Pauli
equation. We formulate the initial condition problem to describe spin-flavor
oscillations in an electromagnetic wave. The transition probabilities
for spin and spin-flavor oscillations are obtained. In case of spin-flavor
oscillations, we analyze the transition and survival probabilities for different
neutrino magnetic moments and various channels of neutrino oscillations. As an application of the obtained results,
we study the possibility of existence of $\nu_{e\mathrm{L}}\to\nu_{\mu\mathrm{R}}$ oscillations
in an electromagnetic wave emitted by a highly magnetized
neutron star. Our results are compared with findings of other authors.
\end{abstract}

\section{Introduction}

The experimental confirmation of oscillations of atmospheric and solar
neutrinos in Refs.~\cite{Fuk98,Ahm02} is the direct indication that neutrinos
have nonzero masses and mixing, which, in its turn, unambiguously
points out to the physics beyond the standard model. This experimental
success was followed by the determination of other parameters in the
neutrino mixing matrix, including $\theta_{13}$ (see, e.g., Ref.~\cite{An12})
and the CP violation phase (see, e.g., Ref.~\cite{Abe17}).

Although the great importance of neutrino flavor oscillations for the
experimental studies of properties of these particles, other channels of 
neutrino oscillations are of interest for the evolution of astrophysical
and cosmological neutrinos~\cite{Raf96}. In the present work, we shall deal mainly
with spin-flavor oscillations of  neutrinos, which imply the conversion
of the type $\nu_{\beta\mathrm{L}}\to\nu_{\alpha\mathrm{R}}$, where
both flavor, $\alpha,\beta = e,\mu,\tau,\dots$, and helicity, $\mathrm{L,R}$, change. This type of neutrino transitions implies
that these particles possess nonzero magnetic moments and interact with
a strong electromagnetic field. Neutrino electromagnetic properties
are reviewed in Ref.~\cite[pp.~461\textendash 479]{FukYan03}.

The majority of studies of neutrino spin-flavor oscillations
involve the neutrino interaction with a constant transverse magnetic
field. Other configurations of electromagnetic fields, including an
electromagnetic wave, were used in Refs.~\cite{EgoLobStu00,DvoStu01}
in the examination of spin-flavor oscillations. The neutrino interaction
with an electromagnetic wave is important for the studies of neutrino
propagation in strong laser pulses~\cite{MeuKeiPia15,Fom18}.

In the present work, we study neutrino spin and spin-flavor oscillations
in a plane electromagnetic wave on the basis of the exact solution
of the Dirac-Pauli equation for a massive neutrino in this external
electromagnetic field~\cite{BagGit90}. In our analysis, we suggest
that neutrinos are Dirac particles. Although some theoretical models
of the neutrino mass generation point out that neutrinos are likely
to be Majorana particles~\cite[pp.~387\textendash 460]{FukYan03}, the issue of the neutrino
nature is still open~\cite{EllFra15}.

Our work is organized as follows. In Sec.~\ref{sec:SPINOSC}, we start with the studies of neutrino
spin oscillations in a plane electromagnetic wave with circular polarization
within one neutrino mass eigenstate. Then, in Sec.~\ref{sec:TMM},
we generalize our treatment to account for neutrino spin-flavor oscillations
with the great transition magnetic moment. The description of spin-flavor
oscillations is based on the formulation of the initial condition
problem for flavor neutrinos~\cite{Dvo11}. The possibility of an astrophysical application
is also considered in Sec.~\ref{sec:TMM}. The influence of the
small diagonal magnetic moments on spin-flavor oscillations is studied in Sec.~\ref{sec:DMM}. Spin-flavor oscillations of neutrinos with great diagonal magnetic moments are considered in Sec.~\ref{sec:GDMM}. We summarize our results in Sec.~\ref{sec:CONCL}. 
The matrix elements of the neutrino spin interaction are calculated in
Appendix~\ref{sec:MATREL}.

\section{Spin oscillations in an electromagnetic wave\label{sec:SPINOSC}}
	
In this section, we consider neutrino spin oscillations, within
one neutrino generation, in a plane electromagnetic wave with the circular
polarization. Our analysis is based on the exact solution of the Dirac-Pauli
equation for a massive neutral fermion found in Ref.~\cite{BagGit90}.

We shall take that a neutrino is a Dirac particle. In this section, we
shall neglect the mixing between different neutrino generations. Assuming
that the considered neutrino mass eigenstate has the nonzero mass
$m$ and the magnetic moment $\mu$, the Dirac equation for such a
neutrino, described by the bispinor $\psi$, in the external electromagnetic
field $F_{\mu\nu}=\partial_{\mu}A_{\nu}-\partial_{\nu}A_{\mu}=(\mathbf{E},\mathbf{B})$
reads
\begin{equation}\label{eq:Direq}
  \left[
    \mathrm{i}\gamma^{\mu}\partial_{\mu}-m-\frac{\mu}{2}\sigma_{\mu\nu}F^{\mu\nu}
  \right]
  \psi=0,
\end{equation}
where $\gamma^{\mu}=(\gamma^{0},\bm{\gamma})$ and $\sigma_{\mu\nu}=\tfrac{\mathrm{i}}{2}\left[\gamma_{\mu},\gamma_{\nu}\right]_{-}$
are the Dirac matrices. In the following, we shall use the standard
representation for the Dirac matrices~\cite{ItzZub80}.

We shall take that the external electromagnetic field is in the form
of a plane electromagnetic wave propagating in the positive direction
of the $z$-axis. It is convenient to choose the following gauge for
the vector potential: $A^{\mu}=(0,\mathbf{A})$. Neglecting the dispersion
of the wave, we can take that $\mathbf{A}=\mathbf{A}(t-z)$. The electric
and magnetic fields are $\mathbf{E}=-\mathbf{A}'$ and $\mathbf{B}=(\mathbf{e}_{z}\times\mathbf{E})$,
where the prime means the derivative with respect to the whole argument
of $\mathbf{A}$.

\begin{figure}
  \centering
  \includegraphics[scale=.3]{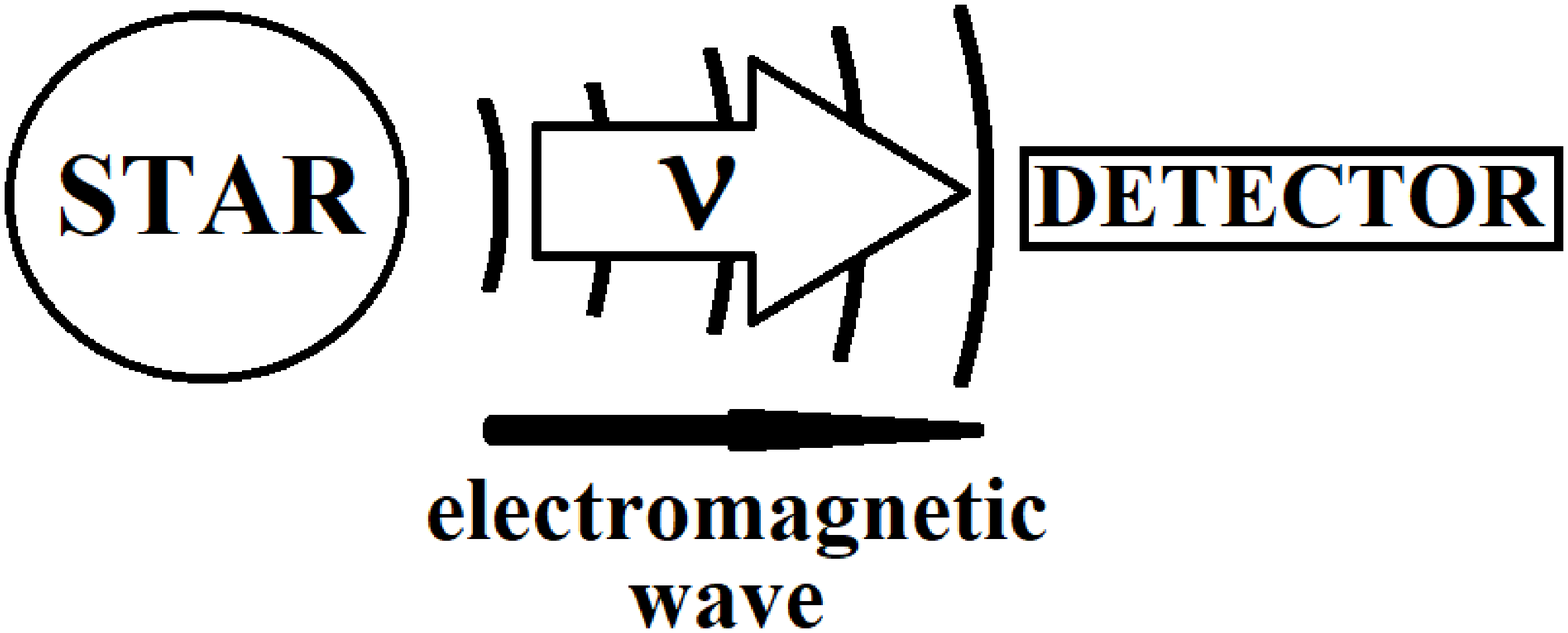}
  \protect
  \caption{The schematic illustration of the neutrino propagation in an electromagnetic wave.
  If both a neutrino and a wave are produced by the same source, e.g., a neutron star
  (see Sec.~\ref{sec:TMM}), they will propagate collinearly towards a detector.
  Thus one studies oscillations of neutrinos, created in a star, under the influence of an 
  electromagnetic wave emitted by the same star.
  It is the motivation to use the neutrino wave function in Eq.~\eqref{eq:psisol}.
  \label{fig:stardet}}
\end{figure}

The exact solution of Eq.~(\ref{eq:Direq}) was found in Ref.~\cite{BagGit90}
for the arbitrary propagation of the fermion with respect to
the electromagnetic wave. In the present work, we consider a special
situation when a neutrino propagates along the wave, i.e. $\psi=\psi(z,t)$.
In this case, the solution of Eq.~(\ref{eq:Direq}) has the form,
\begin{equation}\label{eq:psisol}
  \psi=\frac{e^{-\mathrm{i}Et+\mathrm{i}pz}}{2\sqrt{E(E-p)}}
  \left(
    \begin{matrix}%{c}
      \left[
        m+E-p
      \right]
      v\\
      \left[
        m-E+p
      \right]
      \sigma_{3}v
    \end{matrix}
  \right),
\end{equation}
where $v=v(t-z)$ is the two component spinor, $E=\sqrt{p^{2}+m^{2}}$
is the neutrino energy, $p$ is the neutrino momentum, and $\bm{\sigma}=(\sigma_{1},\sigma_{2},\sigma_{3})$
are the Pauli matrices. It is interesting to notice that $|\psi|^{2}=1$
if $|v|^{2}=1$. The motivation to study neutrinos propagating along the electromagnetic wave results from Fig.~\ref{fig:stardet}.

The spinor $v$ was found in Ref.~\cite{BagGit90} to obey the equation,
\begin{equation}\label{eq:veq}
  \mathrm{i}v'=-\mu(\bm{\sigma}\mathbf{B})v.
\end{equation}
To proceed with the analysis of the solution of Eq.~(\ref{eq:veq})
we suppose that we have a circularly polarized electromagnetic wave,
i.e. $B_{x}=B_{0}\cos\left[\omega(t-z)\right]$, $B_{y}=B_{0}\sin\left[\omega(t-z)\right]$,
and $B_{z}=0$, where $B_{0}$ is the amplitude of the wave and $\omega$
is its frequency. In this case, the solution of Eq.~(\ref{eq:veq})
has the form,
\begin{align}\label{eq:vsol}
  v(\phi)= & \exp(-\mathrm{i}\sigma_{3}\omega\phi/2)
  \left\{
    \cos
    \left[
      \Omega(\phi-\phi_{0})
    \right]+
    \frac{\mathrm{i}}{\Omega}
    \left[
      \mu B_{0}\sigma_{1}+\omega\sigma_{3}/2
    \right]
    \sin
    \left[
      \Omega(\phi-\phi_{0})
    \right]
  \right\}
  \nonumber
  \\
  & \times
  \exp(\mathrm{i}\sigma_{3}\omega\phi_{0}/2)v_{0},
\end{align}
where $\phi=t-z$ is the current phase, $\phi_{0}=t_{0}-z_{0}$ is
the initial phase, $v_{0}=v(\phi_{0})$ is the initial spinor, and
$\Omega=\sqrt{(\mu B_{0})^{2}+\omega^{2}/4}$. One can see in Eq.~(\ref{eq:vsol})
that $v=v_{0}$ if $\phi=\phi_{0}$. Moreover $|v|^{2}=|v_{0}|^{2}$.

Now we should specify the initial condition. Let us suppose that $v_{0}^{\mathrm{T}}=(0,1)$.
In this case, using Eq.~(\ref{eq:psisol}), one can check that $P_{-}\psi_{0}=\psi_{0}$,
where $\psi_{0}$ is the initial bispinor, corresponding to $v=v_0$, $P_{\pm}=(1\pm\Sigma_{z})/2$
are the helicity projection operators, $\bm{\Sigma}=\gamma^{5}\gamma^{0}\bm{\gamma}$
and $\gamma^{5}=\mathrm{i}\gamma^{0}\gamma^{1}\gamma^{2}\gamma^{3}$
are the Dirac matrices. It means that initially a neutrino is left-handed,
i.e. its spin is opposite to the particle momentum.

Neutrino spin oscillations correspond to the appearance of the nonzero component $\psi_{+}$,
i.e. the right-handed polarization, at $\phi>\phi_{0}$, where $\psi_{+}=P_{+}\psi$.
The probability of $\mathrm{L}\to\mathrm{R}$ transitions is $P_{\mathrm{L}\to\mathrm{R}}=\psi^{\dagger}P_{+}\psi$.
Using Eq.~(\ref{eq:psisol}), one can show that $P_{\mathrm{L}\to\mathrm{R}}=\left(1+v^{\dagger}\sigma_{3}v\right)/2$,
i.e. the transition probability is completely defined by the evolution
of the neutrino spin in Eq.~(\ref{eq:veq}).

After lengthy but straightforward calculations based on Eq.~(\ref{eq:vsol}),
we can represent $P_{\mathrm{L}\to\mathrm{R}}$ in the form,
\begin{equation}\label{eq:PLR}
  P_{\mathrm{L}\to\mathrm{R}}=
  \frac{\mu^{2}B_{0}^{2}}{\mu^{2}B_{0}^{2}+\omega^{2}/4}
  \sin^{2}
  \left[
    \sqrt{\mu^{2}B_{0}^{2}+\omega^{2}/4}(\phi-\phi_{0})
  \right].
\end{equation}
One can see in Eq.~(\ref{eq:PLR}) that the transition probability
is a function of $\Delta\phi=\phi-\phi_{0}$, i.e. it depends on both
$t$ and $z$. Thus neutrino spin oscillations can happen both in
time and in space. It is owing to the fact that the spinor $v$ in
Eqs.~(\ref{eq:psisol}) and~(\ref{eq:vsol}) depends on $\phi$.
It is the feature of the field theory based approach for the description
of neutrino oscillations.

The coordinate dependence of $P_{\mathrm{L}\to\mathrm{R}}$ in Eq.~(\ref{eq:PLR})
becomes important if a neutrino interacts with an electromagnetic wave for a
quite long time. In such a situation, the particle initial wave packet becomes broad
enough. It can happen for relatively low energy neutrinos. In practice,
a neutrino is an ultrarelativistic particle. We can localize it and
attribute a mean velocity $\beta=p/E$ of its wave packet. Thus $z=\beta t$
and $z_{0}=\beta t_{0}$.

In this way to modify the coordinate dependence of $P_{\mathrm{L}\to\mathrm{R}}$
in Eq.~(\ref{eq:PLR}), we rederive the transition probability of
neutrino spin oscillations in a circularly polarized electromagnetic
wave obtained in Ref.~\cite{EgoLobStu00} on the basis of the quasiclassical approach. For this purpose one should
set $\Delta m^{2}=0$ and neglect the interaction with matter in the
corresponding expressions in Ref.~\cite{EgoLobStu00} since, in this section, we study
neutrino spin oscillations induced by the electromagnetic wave only
within one neutrino generation. In the following section, we shall
generalize this formalism to describe neutrino spin-flavor oscillations.

\section{Spin-flavor oscillations: great transition magnetic moment\label{sec:TMM}}

In this section, we shall extend the formalism developed in Sec.~\ref{sec:SPINOSC}
to describe neutrino spin-flavor oscillations. For this purpose we
discuss the system of two flavor neutrinos and formulate the initial
condition problem for these particles. Here we will focus on the case of the great
transition magnetic moment. 

For simplicity, we shall consider the system of two flavor neutrinos
$(\nu_{\alpha},\nu_{\beta})$. The generalization of the formalism for greater
number of neutrino flavors is straightforward. For instance, we can
take that $\nu_{\beta}=\nu_{e}$ and $\nu_{\alpha}=\nu_{\mu,\tau}$.
The Lagrangian for this system in the presence of the external electromagnetic
field reads
\begin{equation}\label{eq:Lagrfl}
  \mathcal{L}=\sum_{\lambda=\alpha,\beta}
  \bar{\nu}_{\lambda}\mathrm{i}\gamma^{\mu}\partial_{\mu}\nu_{\lambda} -
  \sum_{\lambda,\lambda'=\alpha,\beta}\bar{\nu}_{\lambda}
  \left(
    m_{\lambda\lambda'}+\frac{1}{2}M_{\lambda\lambda'}\sigma_{\mu\nu}F^{\mu\nu}
  \right)
  \nu_{\lambda'},
\end{equation}
where $(m_{\lambda\lambda'})$ and $(M_{\lambda\lambda'})$ are the
matrices of masses and magnetic moments in the flavor eigenstates
basis. As in Sec.~\ref{sec:SPINOSC}, we shall suppose that only
an electromagnetic wave, propagating along the $z$-axis, is present
in the system and neutrinos move in the direction of the wave. 

We shall define the initial conditions in the system described by
the Lagrangian in Eq.~(\ref{eq:Lagrfl}). For this purpose, we shall suppose that initially $\nu_{\alpha}(z,t=0)=0$ and $\nu_{\beta}(z,t=0)=f\nu_{i\mathrm{L}}$,
where $f=f(z)$ is the given coordinate dependence of the initial
wave function and $\nu_{i\mathrm{L}}$ is a constant bispinor. Let
us suppose that $\Sigma_{z}\nu_{i\mathrm{L}}=-\nu_{i\mathrm{L}}$.
For example, we can take $\nu_{i\mathrm{L}}^{\mathrm{T}}=(1/\sqrt{2})(0,1,0,-1)$.
Such a choice of the initial wave functions corresponds to the presence
of left-handed electron neutrinos (if $\nu_{\beta}\equiv\nu_{e}$)
and the absence of neutrinos of other flavors.

If only two neutrino flavors are present in the system, we can introduce
the neutrino mass eigenstates $\psi_{a}$, $a=1,2$, to diagonalize
the mass matrix in Eq.~(\ref{eq:Lagrfl}). The corresponding matrix
transformation reads
\begin{equation}\label{eq:mattrans}
  \nu_{\lambda}=\sum_{a=1,2}U_{\lambda a}\psi_{a},
  \quad
  (U_{\lambda a})=
  \left(
    \begin{matrix}%{cc}
      \cos\theta & -\sin\theta\\
      \sin\theta & \cos\theta
    \end{matrix}
  \right),
\end{equation}
where $\theta$ is the vacuum mixing angle. Note that we can say whether
neutrinos are Dirac or Majorana only after the transition to the mass
eigenstates basis in Eq.~(\ref{eq:mattrans}).

The Dirac equation for $\psi_{a}$ has the form,
\begin{equation}\label{eq:psiaeq}
  \mathrm{i}\dot{\psi}_{a}=H_{a}\psi_{a}+V\psi_{b},
  \quad
  a,b=1,2,
  \quad
  a\neq b,
\end{equation}
where
\begin{equation}\label{eq:HaV}
  H_{a}=(\bm{\alpha}\mathbf{p})+\beta m_{a}-\mu_{a}\beta
  \left[
    (\bm{\Sigma}\mathbf{B})-\mathrm{i}(\bm{\alpha}\mathbf{E})
  \right],
  \quad
  V=-\mu\beta
  \left[
    (\bm{\Sigma}\mathbf{B})-\mathrm{i}(\bm{\alpha}\mathbf{E})
  \right].
\end{equation}
Here $\mu_{a}\equiv(\mu_{11},\mu_{22})$ and $\mu\equiv\mu_{12}$
are the diagonal and transition magnetic moments in the mass eigenstates
basis, $(\mu_{ab})=U^{\dagger}MU$, as well as $\bm{\alpha}=\gamma^{0}\bm{\gamma}$
and $\beta=\gamma^{0}$ are the Dirac matrices.

First, we shall discuss the situation when $\mu_{a}\ll\mu$. If we neglect
the antiparticle degrees of freedom, the general solution of Eq.~(\ref{eq:psiaeq})
has the form,
\begin{align}\label{eq:psiasol}
  \psi_{a}(z,t) = & \int_{-\infty}^{+\infty}\frac{\mathrm{d}p}{2\pi}
  \sum_{s=\pm}a_{as}e^{-\mathrm{i}E_{a}t+\mathrm{i}pz}u_{as},
  \notag
  \\
  u_{as}= & \frac{1}{2\sqrt{E_{a}(E_{a}-p)}}
  \left(
    \begin{matrix}%{c}
      \left[
        m_{a}+E_{a}-p
      \right]
      v_{s} \\
      \left[
        m_{a}-E_{a}+p
      \right]
      \sigma_{3}v_{s}
    \end{matrix}
  \right),
\end{align}
where $E_a = \sqrt{p^2+m_a^2}$ are the energies of different mass eigenstates, $v_{+}^{\mathrm{T}}=(1,0)$ corresponds to a right-handed neutrino
and $v_{-}^{\mathrm{T}}=(0,1)$ to a left-handed particle since $\sigma_{3}v_{\pm}=\pm v_{\pm}$.
Note that we omit the index $a$ in the spinors $v_{s}$ since we
neglect $\mu_{a}$. We also mention the orthogonality of the basis
bispinors, $u_{a\pm}^{\dagger}u_{a\mp}=0$.

The coefficients $a_{as}$ are supposed to be $c$-number functions
rather than operators acting on Fock states. If we assume that
they depend on neither $t$ nor $z$, the wave functions in Eq.~(\ref{eq:psiasol})
would satisfy the equations $\mathrm{i}\dot{\psi}_{a}=H_{a}\psi_{a}$.
To account for the potential $V$ in Eq.~(\ref{eq:HaV}), which mixes
different mass eigenstates in Eq.~(\ref{eq:psiaeq}), we have to
suppose that $a_{as}$ are no longer constant. Such an approach for
the description of the evolution of neutrino mass eigenstates in the
presence of external fields was used in Ref.~\cite{Dvo11}. However,
unlike Ref.~\cite{Dvo11}, where the corresponding coefficients depend
on time only, here we suppose that $a_{as}=a_{as}(t-z)$. Our main
goal is to study the behavior of these coefficients.

Substituting the ansatz in Eq.~(\ref{eq:psiasol}) to Eq.~(\ref{eq:psiaeq}),
we obtain the equation for $a_{as}$,
\begin{equation}\label{eq:aeqraw}
  \mathrm{i}\sum_{s=\pm}a_{as}'u_{as'}^{\dagger}(1-\alpha_{z})u_{as} =
  \sum_{s=\pm}u_{as'}^{\dagger}Vu_{2s}e^{\mathrm{i}(E_{a}-E_{b})t}a_{bs},
  \quad
  a\neq b.
\end{equation}
Accounting for the mean values
\begin{align}
  u_{as'}^{\dagger}(1-\alpha_{z})u_{as} & =
  \left(
    1 - \frac{p}{E_{a}}
  \right)\delta_{ss'},
  \nonumber
  \\
  u_{1s'}^{\dagger}Vu_{2s} & =
  -\mu v_{s'}^{\dagger}(\bm{\sigma}\mathbf{B})v_{s}
  \sqrt{\frac{(E_{1}-p)(E_{2}-p)}{E_{1}E_{2}}},
  \label{eq:meanu}
  \\
  v_{\pm}^{\dagger}(\bm{\sigma}\mathbf{B})v_{\mp} & =B_{0}\exp[\mp i\omega(t-z)],  
  \label{eq:meanv}
\end{align}
we can rewrite Eq.~(\ref{eq:aeqraw}) in the form,
\begin{equation}\label{eq:aa'}
  \mathrm{i}a_{a\pm}' =-\mu B_{0}\sqrt{\frac{E_{a}(E_{b}-p)}{E_{b}(E_{a}-p)}}
  \exp
  \left[
    \mathrm{i}(E_{a}-E_{b})t\mp\mathrm{i}\omega(t-z)
  \right]
  a_{b\mp}.
\end{equation}
In Eqs.~(\ref{eq:meanu}) and~(\ref{eq:meanv}), we suppose that
one deals with a circularly polarized wave.

Before we proceed with the analysis of the evolution of $a_{as}$
we should specify the initial conditions for them. As in Sec.~\ref{sec:SPINOSC},
we shall suppose that neutrinos are localized along their trajectories.
However, since different mass eigenstates have different masses, we
assume that $z=\bar{\beta}t$, where $\bar{\beta}$ is the center
of inertia velocity~\cite{LanLif94}, $\bar{\beta}=2p/(E_{1}+E_{2})$.
If neutrinos are ultrarelativistic, one gets that $\bar{\beta}\approx1-(m_{1}^{2}+m_{2}^{2})/4p^{2}$.
Therefore, if $t=0$, $a_{as}$ should be taken at zero argument:
$a_{as}(0)$.

Accounting for the initial condition for the flavor eigenstates specified
above, we obtain that
\begin{align}\label{eq:inicondaa}
  a_{1-}(0) = & \sin\theta f(p)
  \left(
    u_{1-}^{\dagger}\nu_{i\mathrm{L}}
  \right),
  \quad
  f(p)=\int_{-\infty}^{+\infty}f(z)e^{-\mathrm{i}pz}\mathrm{d}z
  \notag
  \\
  a_{2-}(0) = & \cos\theta f(p)
  \left(
    u_{2-}^{\dagger}\nu_{i\mathrm{L}}
  \right),
\end{align}
and $a_{a+}(0)=0$. Taking into account the explicit form of the initial
bispinor $\nu_{i\mathrm{L}}$ and $u_{a-}$ in Eq.~(\ref{eq:psiasol}),
we get that $\left(u_{a-}^{\dagger}\nu_{i\mathrm{L}}\right)=m_{a}/\sqrt{2E_{a}(E_{a}-p)}$.
If neutrinos are ultrarelativistic, $\left(u_{a-}^{\dagger}\nu_{i\mathrm{L}}\right)\to1$.

Taking into account the dependence of $a_{as}$ on $t$ in Eq.~(\ref{eq:aa'}),
$a_{as}=a_{as}(t[1-\bar{\beta}])$, we can rewrite Eq.~(\ref{eq:aa'})
as the effective Schr\"odinger equation,
\begin{align}\label{eq:SchPsi}
  \mathrm{i}\frac{\mathrm{d}}{\mathrm{d}t}\Psi= & H_{\mathrm{eff}}\Psi,
  \\
  \notag
  H_{\mathrm{eff}}= & -\mu B_{0}(1-\bar{\beta})
  \left(
    \begin{matrix}%{cccc}
      0 & 0 & 0 & \xi
      \exp
      \left(
        \mathrm{i}\Phi_{-}t
      \right)\\
      0 & 0 & \xi
      \exp
      \left(
        \mathrm{i}\Phi_{+}t
      \right)
      & 0\\
      0 & 
      \exp
      \left(
        -\mathrm{i}\Phi_{+}t
      \right) /
      \xi & 0 & 0\\
      \exp
      \left(
        -\mathrm{i}\Phi_{-}t
      \right) /
      \xi & 0 & 0 & 0
    \end{matrix}
  \right),
\end{align}
where $\Psi^{\mathrm{T}}=\left(a_{1+},a_{1-},a_{2+},a_{2-}\right)$
and
\begin{align}
  \Phi_{\pm} & =E_{1}-E_{2}\pm\omega(1-\bar{\beta})
  \approx
  \frac{m_{1}^{2}-m_{2}^{2}}{2p}\pm\omega\frac{m_{1}^{2}+m_{2}^{2}}{4p^{2}},  
  \label{eq:Phipm}
  \\
  \xi & =\sqrt{\frac{E_{1}(E_{2}-p)}{E_{2}(E_{1}-p)}}
  \approx
  \frac{m_{2}}{m_{1}}.
  \label{eq:xi}
\end{align}
In Eqs.~(\ref{eq:Phipm}) and~(\ref{eq:xi}), we use the approximation
of ultrarelativistic neutrinos.

One can see that $H_{\mathrm{eff}}$ in Eq.~(\ref{eq:SchPsi}) is
non-Hermitian since $\xi\neq1$ because $m_{1}\neq m_{2}$. This fact
results from the assumption that both mass eigenstates with different
masses propagate with the same velocity $\bar{\beta}$. Nevertheless
we can approximately set $\xi=1$. Indeed, let us assume that $m_{a}\sim1\,\text{eV}$~\cite{Ase11}.
Below, we shall be interested in the $\nu_{e}\nu_{\mu}$-oscillations
channel. In this situation, $\delta m^{2}=m_{1}^{2}-m_{2}^{2}\approx7.6\times10^{-5}\,\text{eV}^{2}$~\cite{Abe16}.
Therefore $|m_{1}-m_{2}|/m_{a}\sim10^{-2}$. Hence, we can take that
$\xi=1$ with a sufficient level of accuracy. Note that we should
keep $\delta m^{2}\neq0$ in $\Phi_{\pm}$ in Eq.~(\ref{eq:Phipm})
since these terms contribute to the phase of neutrino oscillations,
which is very sensitive to the change of parameters.

Let us change the variables in Eq.~(\ref{eq:SchPsi}),
\begin{equation}
  \Psi=\mathcal{U}\tilde{\Psi},
  \quad
  \mathcal{U}=\text{diag}
  \left(
    e^{i\Phi_{-}t/2}, e^{i\Phi_{+}t/2},e^{-i\Phi_{+}t/2}, e^{-i\Phi_{-}t/2}
  \right).
\end{equation}
The wave function $\tilde{\Psi}^{\mathrm{T}}=\left(\tilde{a}_{1+},\tilde{a}_{1-},\tilde{a}_{2+},\tilde{a}_{2-}\right)$
obeys the modified Schr\"odinger equation,
\begin{align}\label{eq:SchPsitilde}
  \mathrm{i}\frac{\mathrm{d}}{\mathrm{d}t}\tilde{\Psi}= & 
  \tilde{H}_{\mathrm{eff}}\tilde{\Psi},
  \notag
  \\
  \tilde{H}_{\mathrm{eff}}= &
  \left(
    \begin{matrix}%{cccc}
      \Phi_{-}/2 & 0 & 0 & -\mu B_{0}(1-\bar{\beta})\\
      0 & \Phi_{+}/2 & -\mu B_{0}(1-\bar{\beta}) & 0\\
      0 & -\mu B_{0}(1-\bar{\beta}) & -\Phi_{+}/2 & 0\\
      -\mu B_{0}(1-\bar{\beta}) & 0 & 0 & -\Phi_{-}/2
    \end{matrix}
  \right).
\end{align}
We recall that we set $\xi=1$ in the $\mu B_{0}$-terms in Eq.~(\ref{eq:SchPsitilde}).

The solution of Eq.~(\ref{eq:SchPsitilde}) has the form,
\begin{equation}\label{eq:tildePsisol}
  \tilde{\Psi}(t)=\sum_{\zeta=\pm}
  \left[
    e^{-i\Omega_{\zeta}t}
    \left(
      U_{\zeta}\otimes U_{\zeta}^{\dagger}
    \right)+
    e^{i\Omega_{\zeta}t}
    \left(
      V_{\zeta}\otimes V_{\zeta}^{\dagger}
    \right)
  \right]
  \tilde{\Psi}_{0},
\end{equation}
where $\Omega_{\pm}=\sqrt{\mu^{2}B_{0}^{2}(1-\bar{\beta})^{2}+\Phi_{\pm}^{2}/4}$
and
\begin{align}\label{eq:UV}
  U_{+} & = 
  \frac{\sqrt{\Omega_{+}+\Phi_{+}/2}}{\sqrt{2\Omega_{+}}}
  \left(
    \begin{matrix}%{c}
      0\\
      1\\
      -\dfrac{\mu B_{0}(1-\bar{\beta})}{\Omega_{+}+\Phi_{+}/2}\\
      0
    \end{matrix}
  \right),
  \quad
  V_{+}=\frac{\sqrt{\Omega_{+}+\Phi_{+}/2}}{\sqrt{2\Omega_{+}}}
  \left(
    \begin{matrix}%{c}
      0\\
      \dfrac{\mu B_{0}(1-\bar{\beta})}{\Omega_{+}+\Phi_{+}/2}\\
      1\\
      0
    \end{matrix}
  \right),
  \nonumber
  \\
  U_{-} & =\frac{\sqrt{\Omega_{-}+\Phi_{-}/2}}{\sqrt{2\Omega_{-}}}
  \left(
    \begin{matrix}%{c}
      1\\
      0\\
      0\\
      -\dfrac{\mu B_{0}(1-\bar{\beta})}{\Omega_{-}+\Phi_{-}/2}
    \end{matrix}
  \right),
  \quad
  V_{-}=\frac{\sqrt{\Omega_{-}+\Phi_{-}/2}}{\sqrt{2\Omega_{-}}}
  \left(
    \begin{matrix}%{c}
      \dfrac{\mu B_{0}(1-\bar{\beta})}{\Omega_{-}+\Phi_{-}/2}\\
      0\\
      0\\
      1
    \end{matrix}
  \right),
\end{align}
are the eigenvectors of $\tilde{H}_{\mathrm{eff}}$: $\tilde{H}_{\mathrm{eff}}U_{\zeta}=\Omega_{\zeta}U_{\zeta}$
and $\tilde{H}_{\mathrm{eff}}V_{\zeta}=-\Omega_{\zeta}V_{\zeta}$.

Using Eqs.~(\ref{eq:inicondaa}), (\ref{eq:tildePsisol}) and~(\ref{eq:UV}),
we find that the coefficients $a_{a+}(t)$ have the form,
\begin{align}\label{eq:a12p}
  a_{1+}(t) & =\mathrm{i}e^{\mathrm{i}\Phi_{-}t/2}
  (1-\bar{\beta})\frac{\mu B_{0}}{\Omega_{-}}\sin(\Omega_{-}t)f(p)\cos\theta,
  \nonumber
  \\
  a_{2+}(t) & =\mathrm{i}e^{-\mathrm{i}\Phi_{+}t/2}
  (1-\bar{\beta})\frac{\mu B_{0}}{\Omega_{+}}\sin(\Omega_{+}t)f(p)\sin\theta.
\end{align}
The explicit form of $a_{a-}(t)$ is not important for our purposes
since these coefficients do not contribute to the evolution of the right-handed neutrino
states.

We are interested in the appearance of right-handed neutrinos of the
flavor $\alpha$ in a beam initially consisting of $\nu_{\beta\mathrm{L}}$.
We shall suppose that the initial wave packet is quite wide, i.e.
we take that $f(p)=2\pi\delta(p-p_{0})$, where $p_{0}$ is the initial
momentum. In the following, we shall omit the subscript 0 for brevity.
Using Eqs.~(\ref{eq:mattrans}), (\ref{eq:psiasol}), and~(\ref{eq:a12p}),
we obtain that the wave function $\nu_{\alpha\mathrm{R}}$ has the
form,
\begin{align}\label{eq:nualphaR}
  \nu_{\alpha\mathrm{R}}(z,t)= &
  \int_{-\infty}^{+\infty}\frac{\mathrm{d}p}{2\pi}
  \left[
    \cos\theta e^{-\mathrm{i}E_{1}t}a_{1+}(t)u_{1+}-
    \sin\theta e^{-\mathrm{i}E_{2}t}a_{2+}(t)u_{2+}
  \right]
  \nonumber
  \\
  & =
  \left[
    \cos\theta e^{-\mathrm{i}E_{1}t}a_{1+}(t)
    \left(
      \nu_{f\mathrm{R}}^{\dagger}u_{1+}
    \right) -
    \sin\theta e^{-\mathrm{i}E_{2}t}a_{2+}(t)
    \left(
      \nu_{f\mathrm{R}}^{\dagger}u_{2+}
    \right)
  \right]
  \nu_{f\mathrm{R}},
\end{align}
where $\nu_{f\mathrm{R}}^\mathrm{T}=(1/\sqrt{2})(1,0,1,0)$. Basing on Eq.~(\ref{eq:psiasol}),
one gets that $\left(\nu_{f\mathrm{R}}^{\dagger}u_{a+}\right)= m_{a} [2E_{a}(E_{a}-p)]^{-1/2}$.
If we study ultrarelativistic neutrinos, then $\left(\nu_{f\mathrm{R}}^{\dagger}u_{a+}\right)\to1$.

Using Eqs.~(\ref{eq:a12p}) and~(\ref{eq:nualphaR}), as well as
considering ultrarelativistic neutrinos, we obtain the transition
probability for $\nu_{\beta\mathrm{L}}\to\nu_{\alpha\mathrm{R}}$
oscillations,
\begin{equation}\label{eq:PbetaLalphaR}
  P_{\nu_{\beta\mathrm{L}}\to\nu_{\alpha\mathrm{R}}}(t)=|\nu_{\alpha \mathrm{R}}|^{2}=
  \mu^{2}B_{0}^{2}(1-\bar{\beta})^{2}
  \left[
    \cos^{2}\theta\frac{\sin(\Omega_{-}t)}{\Omega_{-}}-
    \sin^{2}\theta\frac{\sin(\Omega_{+}t)}{\Omega_{+}}
  \right]^{2}.
\end{equation}
It is interesting to compare $P_{\nu_{\beta\mathrm{L}}\to\nu_{\alpha\mathrm{R}}}$
in Eq.~(\ref{eq:PbetaLalphaR}) with the transition probability of
spin-flavor oscillations of Dirac neutrinos with large transition
magnetic moment in a constant transverse magnetic field studied in
Ref.~\cite{DvoMaa07}. For this purpose we should set $\omega=0$
and replace $(1-\bar{\beta})\to1$ in Eq.~(\ref{eq:PbetaLalphaR}).
In this case, $\Omega_{+}=\Omega_{-}=\sqrt{(\mu B_{0})^{2}+(\delta m^{2}/4p)^{2}}$.
Thus we reproduce the transition probability found in Ref.~\cite{DvoMaa07}.

Now we can compare $P_{\nu_{\beta\mathrm{L}}\to\nu_{\alpha\mathrm{R}}}$
in Eq.~(\ref{eq:PbetaLalphaR}) with the corresponding transition
probability for neutrino spin-flavor oscillations in a plane electromagnetic
wave found in Ref.~\cite{EgoLobStu00}.
Using Eq.~(21) in Ref.~\cite{EgoLobStu00} and omitting the neutrino matter interaction there, since here we study the neutrino interaction with an electromagnetic wave only, one obtains the following transition probability for $\nu_{\beta\mathrm{L}}\to\nu_{\alpha\mathrm{R}}$ oscillations:
\begin{align}\label{eq:Pwrong}
  P_{\nu_{\beta\mathrm{L}}\to\nu_{\alpha\mathrm{R}}}(t) = &
  \frac{\mu^{2}B_{0}^{2}(1-\bar{\beta})^{2}}
  {\mu^{2}B_{0}^{2}(1-\bar{\beta})^{2} +
  [\delta m^2 A(\theta)/4p \pm \omega(1-\bar{\beta})/2]^2}
  \notag
  \\
   & \times
  \sin^2
  \left(
    \sqrt{\mu^{2}B_{0}^{2}(1-\bar{\beta})^{2} +
    [\delta m^2 A(\theta)/4p \pm \omega(1-\bar{\beta})/2]^2} t
  \right).
\end{align}
If we consider $\nu_{e\mathrm{L}}\to\nu_{\mu\mathrm{R}}$ transitions (see below), the function $A(\theta)$ reads $A(\theta) = (1+\cos2\theta)/2$~\cite{LikStu95}. The signs $\pm$ in Eq.~\eqref{eq:Pwrong} correspond to different polarizations of the wave.

One can see that the result
of Ref.~\cite{EgoLobStu00}, shown in Eq.~\eqref{eq:Pwrong}, cannot be reproduced by Eq.~(\ref{eq:PbetaLalphaR}), except in the trivial case $\theta=0$.
However, the situation of a zero mixing angle between active neutrinos was recently excluded experimentally in Ref.~\cite{An12}, where it was shown that the remaining mixing angle $\theta_{13}$ is nonzero.
Moreover, basing on Eq.~(\ref{eq:PbetaLalphaR}), one cannot expect
the appearance of the resonant amplification of spin-flavor oscillations, which can result from Eq.~\eqref{eq:Pwrong} and was claimed in Ref.~\cite{EgoLobStu00}.

The mentioned discrepancy is because of the unsubstantiated account
of the neutrino vacuum oscillations phase $\delta m^{2}A(\theta)/4p$ in the
dynamics of the neutrino spin in Ref.~\cite{EgoLobStu00}. In Sec.~\ref{sec:SPINOSC},
using the method of the exact solution of the Dirac-Pauli equation, we
have confirmed that the results of Ref.~\cite{EgoLobStu00} are applicable
for the description of the neutrino spin evolution within one mass
eigenstate. Thus, in the present work, we have generalized the findings of Ref.~\cite{EgoLobStu00}
to correctly treat neutrino spin-flavor oscillations.

To analyze Eq.~(\ref{eq:PbetaLalphaR}) we suppose that $\delta\Omega\ll\Omega_{\pm}$,
where $\delta\Omega=\left(\Omega_{+}-\Omega_{-}\right)/2$. In this
situation, the function $P_{\nu_{\beta\mathrm{L}}\to\nu_{\alpha\mathrm{R}}}$
becomes rapidly oscillating, being modulated by a slowly varying envelope
function. Therefore we can apply the method of the analysis of this
envelope function developed in Refs.~\cite{Dvo08,DvoMaa09}. Let
us represent $P_{\nu_{\beta\mathrm{L}}\to\nu_{\alpha\mathrm{R}}}$
in the form,
\begin{equation}\label{eq:PLRenv}
  P_{\nu_{\beta\mathrm{L}}\to\nu_{\alpha\mathrm{R}}}(t) =
  A_{\mathrm{eff}}(t)\sin^{2}(\Omega_{0}t),
  \quad
  A_{\mathrm{eff}}(t)=A_{\mathrm{min}}+2\delta A\sin^{2}(\delta\Omega t),
\end{equation}
where
\begin{equation}\label{eq:AmindA}
  A_{\mathrm{min}}=
  \frac{
  \left[
    \mu B_{0}(1-\bar{\beta})\cos2\theta
  \right]^{2}
  }{\Omega_{0}^{2}}
  \left(
    1+\frac{2\delta\Omega}{\cos2\theta\Omega_{0}}
  \right),
  \quad
  \delta A=
  \frac{
  \left[
    \mu B_{0}(1-\bar{\beta})\sin2\theta
  \right]^{2}
  }{2\Omega_{0}^{2}},
\end{equation}
are the minimal value and the amplitude of the envelope function $A_{\mathrm{eff}}(t)$.
In Eqs.~(\ref{eq:PLRenv}) and~(\ref{eq:AmindA}), $\Omega_{0}=\sqrt{\mu^{2}B_{0}^{2}(1-\bar{\beta})^{2}+(\delta m^{2}/4p)^{2}}$.

\begin{figure}
  \centering
  \subfigure[]
  {\label{1a}
  \includegraphics[scale=.35]{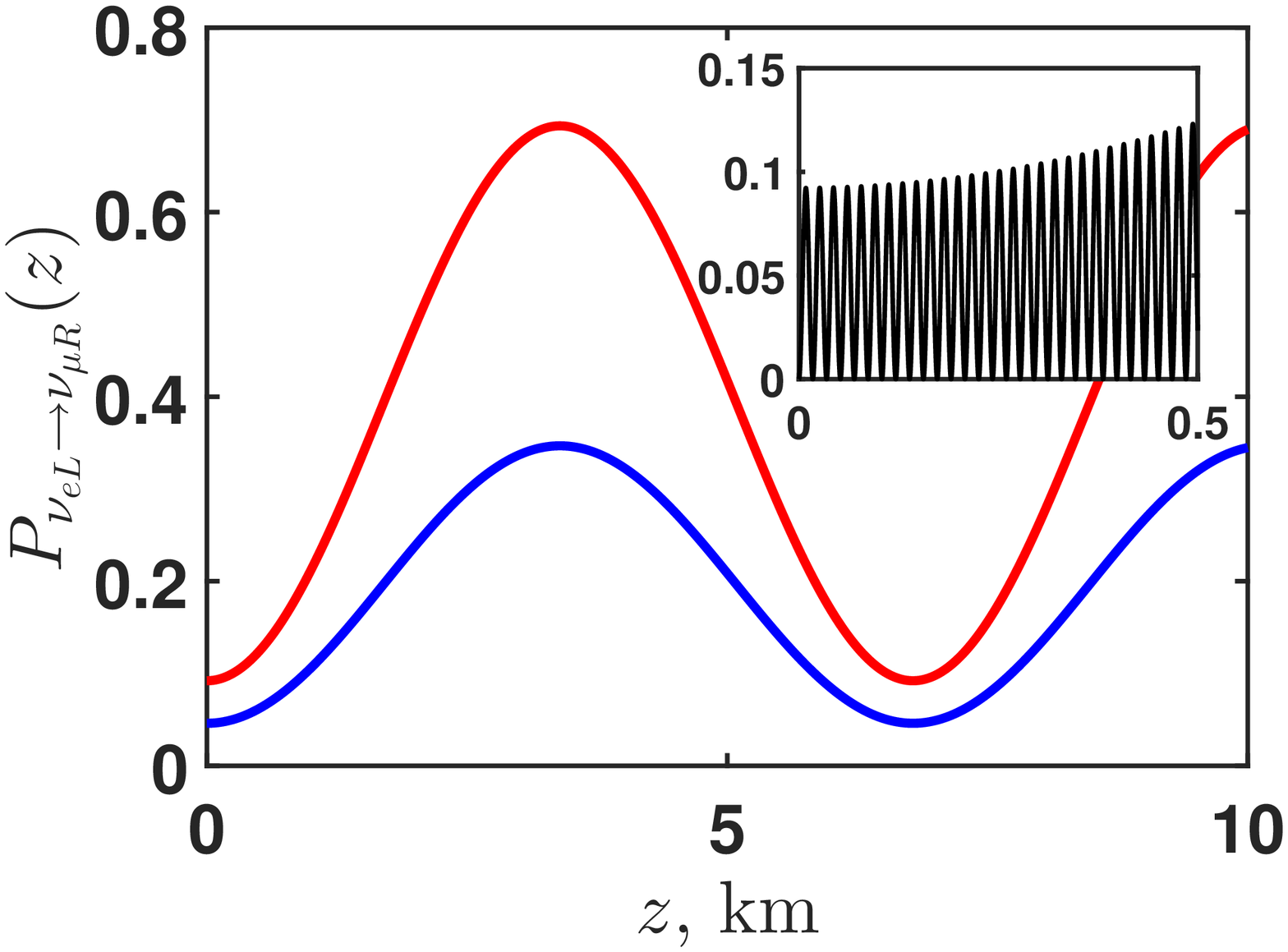}}
  \hskip-.6cm
  \subfigure[]
  {\label{1b}
  \includegraphics[scale=.35]{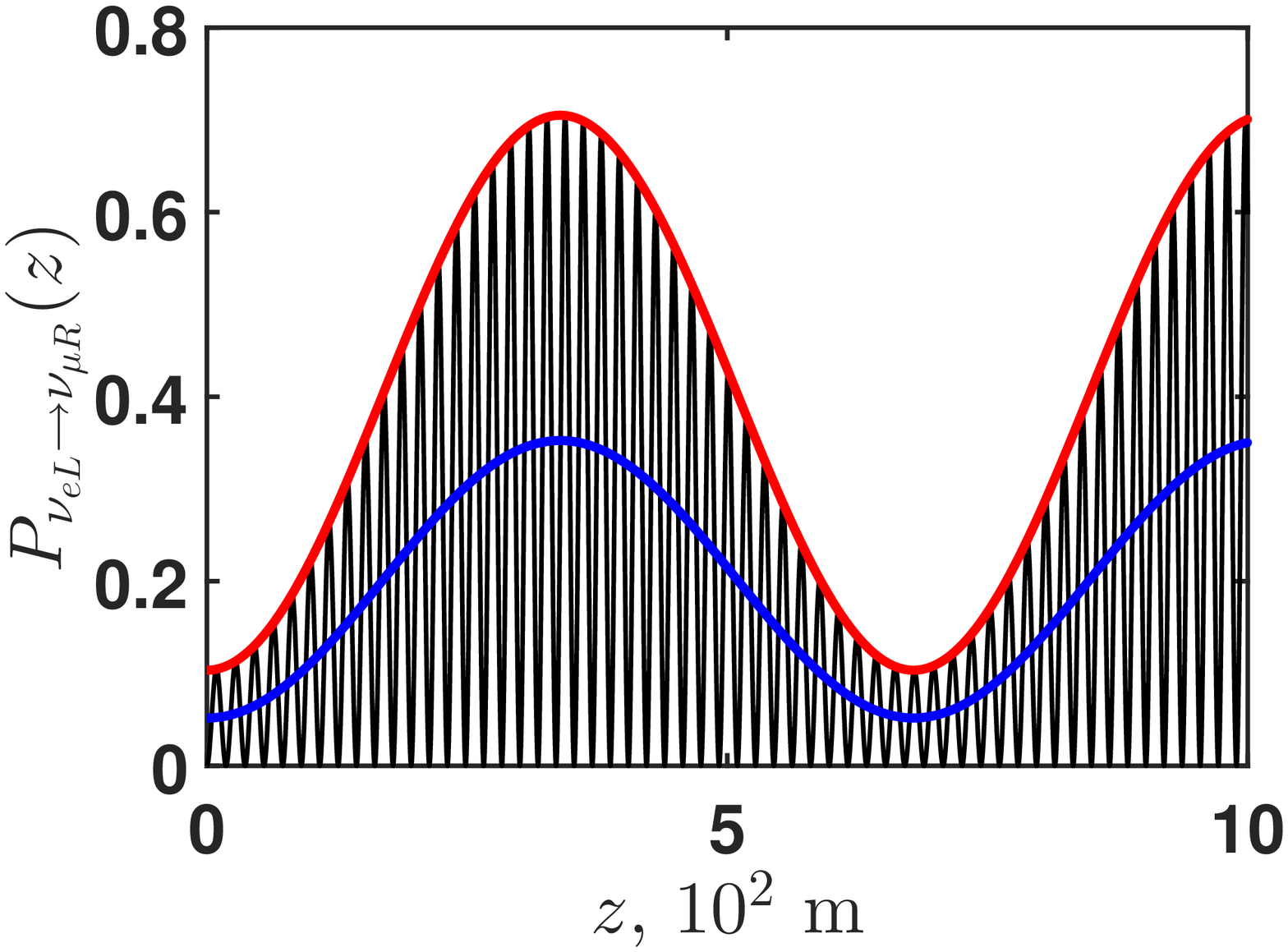}}
  \protect
  \caption{The transition probability for the $\nu_{e\mathrm{L}}\to\nu_{\mu\mathrm{R}}$
  oscillations channel for different $\omega$ versus the propagation
  distance $z=t$, based on Eq.~(\ref{eq:PLRenv}). The neutrino parameters
  are $p=1\,\text{keV}$, $m_{1}\approx m_{2}=1\,\text{eV}$,
  $\delta m^{2}=7.6\times10^{-5}\,\text{eV}^{2}$,
  $\theta=0.6$~\cite{Abe16}, and $\mu=10^{-11}\mu_{\mathrm{B}}$~\cite{Bed13},
  where $\mu_{\mathrm{B}}$ is the Bohr magneton. The amplitude of the
  electromagnetic wave $B_{0}=10^{18}\,\text{G}$. 
  (a)~$\omega=10^{12}\,\text{s}^{-1}$;
  (b)~$\omega=10^{13}\,\text{s}^{-1}$.
  The envelope function $A_{\mathrm{eff}}(t)$
  and the averaged transition probability $\bar{P}(t)$ are depicted
  by the red and blue lines respectively.
  We do not show the actual behavior of the transition probability in panel~(a) since it oscillates 
  very rapidly. Instead we represent it in the inset in panel~(a) at the small propagation
  distance $z<500\,\text{m}$.\label{fig:PTMM}}
\end{figure}

The behavior of the transition probability for the $\nu_{e\mathrm{L}}\to\nu_{\mu\mathrm{R}}$
oscillations channel for different $\omega$ is shown in Fig.~\ref{fig:PTMM}.
One can see that $P_{\nu_{e\mathrm{L}}\to\nu_{\mu\mathrm{R}}}$ is
a rapidly oscillating function. Therefore the averaged signal $\bar{P}(t)=A_{\mathrm{min}}/2+\delta A\sin^{2}(\delta\Omega t)$
will be detected. The maximal averaged transition probability reads
\begin{equation}\label{eq:Pavmax}
  \bar{P}_{\mathrm{max}}=
  \frac{
  \left[
    \mu B_{0}(1-\bar{\beta})
  \right]^{2}
  }{2\Omega_{0}^{2}}
  \left(
    1+2\cos2\theta\frac{\delta\Omega}{\Omega_{0}}
  \right).
\end{equation}
The envelope function $A_{\mathrm{eff}}(t)$ and the averaged transition
probability $\bar{P}(t)$ are also shown in Fig.~\ref{fig:PTMM}.

The maximal averaged transition probability, calculated using Eq.~(\ref{eq:Pavmax})
with the parameters corresponding to Figs.~\ref{1a} and~\ref{1b}, is $\bar{P}_{\mathrm{max}}\approx0.35$.
This result is in the agreement with Figs.~\ref{1a} and~\ref{1b}. We can also estimate
the modulation length $L=\pi/\delta\Omega$. Using Eq.~(\ref{eq:PLRenv}),
one gets that
\begin{equation}\label{eq:dOmega}
  \delta\Omega=\frac{\omega\delta m^{2}}{32\Omega_{0}p^{3}}
  \left(
    m_{1}^{2}+m_{2}^{2}
  \right).
\end{equation}
Basing on the parameters corresponding to Fig.~\ref{1a} and Eq.~(\ref{eq:dOmega}),
we obtain that $L\approx7.2\,\text{km}$, whereas $L\approx720\,\text{m}$ for Fig.~\ref{1b}.  These values of $L$ are in the agreement with Fig.~\ref{fig:PTMM}. 

As an application of the obtained results, we shall consider the possibility of $\nu_{e\mathrm{L}}\to\nu_{\mu\mathrm{R}}$
spin-flavor oscillations of astrophysical neutrinos in the vicinity
of a neutron star (NS). A magnetic field with the strength $\sim10^{18}\,\text{G}$,
chosen in Fig.~\ref{fig:PTMM}, is slightly weaker than the strongest
field allowed in NS~\cite{CarPraLat01}. The electromagnetic emission of pulsars ranges from radio to X-ray wavelengths~\cite{Man04}.
The frequencies adopted in Fig.~\ref{fig:PTMM} correspond to wavelengths $\lambda = 1.8\,\text{mm}$ in Fig.~\ref{1a} and $\lambda = 0.18\,\text{mm}$ in Fig.~\ref{1b}. These wavelengths lie in the infrared region. The electromagnetic emission of NS with such wavelengths was described in Ref.~\cite{Dan11}.

Note that further enhancement of the frequency is inexpedient since $\delta\Omega$ becomes comparable with $\Omega_{\pm}$ then. In principle, using our approach, we can also analyze spin-flavor oscillations in electromagnetic waves with wavelengths in the range: $(\text{several mm})<\lambda<(\text{several tens of cm})$, emitted by radio pulsars. For a radio pulsar, $L$ is greater than that shown in Fig.~\ref{fig:PTMM}.

%The frequency $\sim10^{3}\,\text{s}^{-1}$,
%used in Fig.~\ref{1a}, can be associated to the rotation frequency
%of a millisecond pulsar. Therefore an electromagnetic wave with such
%characteristics ($B_{0}=10^{18}\,\text{G}$ and $\omega=10^{3}\,\text{s}^{-1}$)
%can be emitted by a rapidly rotating highly magnetized NS.

%The half of the modulation length, $L/2$, corresponds to the maximal
%conversion of the initial neutrino beam. Using the averaged transition
%probability in Fig.~\ref{1a}, one finds that $L/2\sim3.6\,\text{\text{km}}$, whereas $L/2\sim360\,\text{\text{m}}$ in Fig.~\ref{1b}.
An electromagnetic wave, which causes neutrino spin-flavor oscillations, can be considered as plane only in the near field zone. In the far field zone, a wave should be taken as spherical, with its amplitude decreasing as $1/z$. The size of the near field zone was estimated in Ref.~\cite{Rap02} as $l \sim D^2/\lambda \sim D^2\omega$, where $D$ is the length scale of the region where a wave is emitted. In case of a pulsar, we can take that $D \sim 10^4\,\text{cm}$~\cite{Bau15}. Thus, we get that $l \sim 3\times 10^4\,\text{km}$ for the case shown in Fig.~\ref{1a}, and $l \sim 3\times 10^5\,\text{km}$ for Fig.~\ref{1b}. The obtained values of $l$ are much greater than $L$ in Fig.~\ref{fig:PTMM}. Hence the approximation of a plane electromagnetic wave is valid.

We studied neutrino spin-flavor oscillations in constant in space magnetic fields in Refs.~\cite{DvoMaa07,Dvo11}. Although neutrino oscillations in a constant field are not suppressed by the factor $(1-\bar{\beta}) \ll 1$, cf. Eq.~\eqref{eq:PbetaLalphaR}, the analytical expressions for the transition probabilities, derived in Refs.~\cite{DvoMaa07,Dvo11}, cannot be applied for neutrinos propagating near NS at distances $R_\mathrm{NS} < z \lesssim l$, where $R_\mathrm{NS}\sim10\,\text{km}$ is the typical NS radius. Indeed, one cannot neglect the coordinate dependence of the magnetic field in NS at such distances, i.e. we have to take  $B(z) \sim B_\text{surf} (R/z)^3$, where $B_\text{surf}$ is the magnetic field at the NS surface. Thus only a numerical transition probability 
can be obtained by solving the effective Schr\"odinger equation with $B=B(z)$. On the contrary, the analytical expression in Eq.~\eqref{eq:PbetaLalphaR} is valid for the distances $z$ up to $l$.

The most powerful neutrino emission takes place during a supernova
explosion. In several minutes after an explosion, the neutrino flux
becomes rather weak to be detected. Nevertheless, when the NS temperature is above
$10^{7}\,\text{K}\sim1\,\text{keV}$, neutrinos carry away energy
from NS and are an effective tool for the NS cooling~\cite{HaePotYak07}.
That is why we take the neutrino energy $\sim1\,\text{keV}$ in Fig.~\ref{fig:PTMM}.
Thus such neutrinos, produced in modified Urca processes in the NS
core, are subject to the described spin-flavor oscillations. Basing
on $\bar{P}_{\mathrm{max}}$ in Fig.~\ref{fig:PTMM}, one gets that about
$35\%$ of $\nu_{e\mathrm{L}}$ will be converted to $\nu_{\mu\mathrm{R}}$.

It should be noted that the chosen parameters $B_{0}$ and $p$ correspond
to $\delta A\lesssim1$ in Eq.~(\ref{eq:AmindA}). It guarantees
that the transition probability is not suppressed significantly. We mention that the study of the electromagnetic emission of pulsars is a rapidly developing area of modern astrophysics. Although the recent advances in the description of the pulsars radiation~\cite{Bes17}, there is no self-consistent model of this process~\cite{Har17}. In realistic situation, the amplitude of an electromagnetic wave is likely to be smaller that $10^{18}\,\text{G}$. In the case, when $\mu B_{0}(1-\bar{\beta}) \ll \delta m^{2}/4p$, using Eq.~\eqref{eq:Pavmax}, the maximal averaged transition probability is
\begin{equation}\label{eq:PtrsmallB}
  \bar{P}_\mathrm{max} \approx 
  2
  \left[
    \mu B_{0}(1-\bar{\beta}) \frac{2 p}{\delta m^2}
  \right]^{2} < 0.35.
\end{equation}
Thus, the value $\bar{P}_\mathrm{max} = 0.35$, obtained above, is the upper bound on the maximal transition probability in NS since we supposed that $B_0$ equals to the strongest magnetic field, which can exist in such a system.

%Figure~\ref{1b} is unlikely to be implemented in astrophysical environments
%because of a huge frequency of the electromagnetic wave $\omega=10^{13}\,\text{s}^{-1}$.
%It is provided to be compared with Fig.~\ref{1a} to demonstrate the
%dependence of $P_{\nu_{e\mathrm{L}}\to\nu_{\mu\mathrm{R}}}$ on $\omega$.

At the end of this section, we discuss the general issue on the approximations
used in the derivation of the main results. To obtain Eq.~(\ref{eq:inicondaa})
we supposed that the neutrino mass eigenstates are localized on their classical trajectories. However, to derive Eq.~(\ref{eq:nualphaR})
one has to assume that the neutrino wave packets are wide. The latter
assumption is necessary for a significant overlap of different
mass eigenstates for the oscillations process to occur. Formally we should
require that $\ell\gg\delta\Omega^{-1}$, where $\ell$ is the
width of the wave packet. However we can suggest that point-like neutrinos
with equal velocities are coherently emitted during the time interval
$\Delta t\gg\delta\Omega^{-1}$. In this case, one deals with an effectively
wide neutrino wave packet consisting of mass eigenstates propagating
with a certain velocity. Thus we can reconcile both assumptions necessary
to get Eqs.~(\ref{eq:inicondaa}) and~(\ref{eq:nualphaR}).

\section{Spin-flavor oscillations: influence of diagonal magnetic moments\label{sec:DMM}}

In this section, we discuss the influence of the diagonal magnetic
moments $\mu_{a}$ on the neutrino spin-flavor oscillations in a plane
electromagnetic wave. In our analysis, we shall still suppose that $\mu_{a}$ is small: $\mu_{a}\ll\mu$.

If one accounts for $\mu_{a}$, the main formalism, developed in Sec.~\ref{sec:TMM},
remains practically unchanged. First, Eq.~(\ref{eq:psiasol}) is
modified, so that the two component spinor $v_{s}$, entering to the bispinor $u_{as}$, becomes dependent on time and
$\mu_{a}$, i.e. $v_{s}\to v_{as}(t)$. The temporal dependence of
$v_{as}$ is given by Eq.~(\ref{eq:vsol}). We can choose two independent
spin states $v_{0s}$ in Eq.~(\ref{eq:vsol}) as $v_{0+}^{\mathrm{T}}=(1,0)$
and $v_{0-}^{\mathrm{T}}=(0,1)$. In this situation, both $v_{a\pm}(t)$
and $u_{a\pm}$, corresponding to the opposite spin states, are orthogonal. 

Second, the matrix elements in Eq.~(\ref{eq:meanv}) become quite
cumbersome. The calculation of the matrix elements is provided in
Appendix~\ref{sec:MATREL} in the general form. Using Eq.~(\ref{eq:veps})
and making the calculations similar to those in Sec.~\ref{sec:TMM},
we derive the analogue of Eq.~(\ref{eq:SchPsi}),
\begin{align}\label{eq:SchPsiDMM}
  \mathrm{i}\frac{\mathrm{d}}{\mathrm{d}t}\Psi = & H_{\mathrm{eff}}\Psi,
  \notag
  \\
  H_{\mathrm{eff}}= & -\mu B_{0}(1-\bar{\beta})
  \left(
    \begin{matrix}%{cccc}
      0 & 0 & \epsilon_{+} &
      \exp
      \left(
        \mathrm{i}\Phi_{-}t
      \right) \\
      0 & 0 &
      \exp
      \left(
        \mathrm{i}\Phi_{+}t
      \right) &
      \epsilon_{-}\\
      \epsilon_{+}^{*} &
      \exp
      \left(
        -\mathrm{i}\Phi_{+}t
      \right) &
      0 & 0\\
      \exp
      \left(
        -\mathrm{i}\Phi_{-}t
      \right) &
      \epsilon_{-}^{*} & 0 & 0
    \end{matrix}
  \right),
\end{align}
where
\begin{equation}\label{eq:epspm}
  \epsilon_{\pm}=\frac{B_{0}}{\omega}
  \left[
    \pm
    \left(
      \mu_{1}+\mu_{2}
    \right)
    \mp\mu_{1}e^{\pm \mathrm{i}\omega t(1-\bar{\beta})}
    \mp\mu_{2}e^{\mp \mathrm{i}\omega t(1-\bar{\beta})}
  \right].
\end{equation}
In Eqs.~(\ref{eq:SchPsiDMM}) and~(\ref{eq:epspm}), we set $z=\bar{\beta}t$
and assume that $m_{1}\approx m_{2}$ in the off-diagonal terms $\sim\exp(\pm\mathrm{i}\Phi_{\pm}t)$
in $H_{\mathrm{eff}}$.

Since we discuss the situation when $m_{1}\approx m_{2}$, it is reasonable
to take that $\mu_{1}\approx\mu_{2}=\mu'$.
%, since $\mu_{a}\sim m_{a}$; cf. Ref.~\cite{DvoStu04}.
In this case, $\epsilon_{\pm}=\pm\epsilon$,
where $\epsilon=4\mu'B_{0}\sin^{2}\left[\omega t(1-\bar{\beta})/2\right]/\omega$.
Then, we can derive the analogue of Eq.~(\ref{eq:SchPsitilde}),
\begin{align}\label{eq:SchPsitildeDMM}
  \mathrm{i}\frac{\mathrm{d}}{\mathrm{d}t}\tilde{\Psi}= &
  \tilde{H}_{\mathrm{eff}}\tilde{\Psi},
  \notag
  \\
  \tilde{H}_{\mathrm{eff}}= &
  \left(
    \begin{matrix}%{cccc}
      \Phi_{-}/2 & 0 & -\mu B_{0}(1-\bar{\beta})\epsilon & -\mu B_{0}(1-\bar{\beta})\\
      0 & \Phi_{+}/2 & -\mu B_{0}(1-\bar{\beta}) & \mu B_{0}(1-\bar{\beta})\epsilon\\
      -\mu B_{0}(1-\bar{\beta})\epsilon & -\mu B_{0}(1-\bar{\beta}) & -\Phi_{+}/2 & 0\\
      -\mu B_{0}(1-\bar{\beta}) & \mu B_{0}(1-\bar{\beta})\epsilon & 0 & -\Phi_{-}/2
    \end{matrix}
  \right).
\end{align}
Should one have the solution of Eq.~(\ref{eq:SchPsitildeDMM}), the
transition probability for $\nu_{\beta\mathrm{L}}\to\nu_{\alpha\mathrm{R}}$
oscillations reads
\begin{equation}\label{eq:PLRtildea}
  P_{\nu_{\beta\mathrm{L}}\to\nu_{\alpha\mathrm{R}}}(t)=
  |\cos\theta\tilde{a}_{1+}(t)-\sin\theta\tilde{a}_{2+}(t)|^{2}.
\end{equation}
One can show that Eq.~(\ref{eq:PLRtildea}) is equivalent to $P_{\nu_{\beta\mathrm{L}}\to\nu_{\alpha\mathrm{R}}}$
calculated using $a_{a+}(t)$, i.e. based on the solution of Eq.~(\ref{eq:SchPsiDMM}),
as $\nu_{\alpha\mathrm{R}}(z,t)$ in Eq.~(\ref{eq:nualphaR}).

\begin{figure}
  \centering
  \subfigure[]
  {\label{2a}
  \includegraphics[scale=.35]{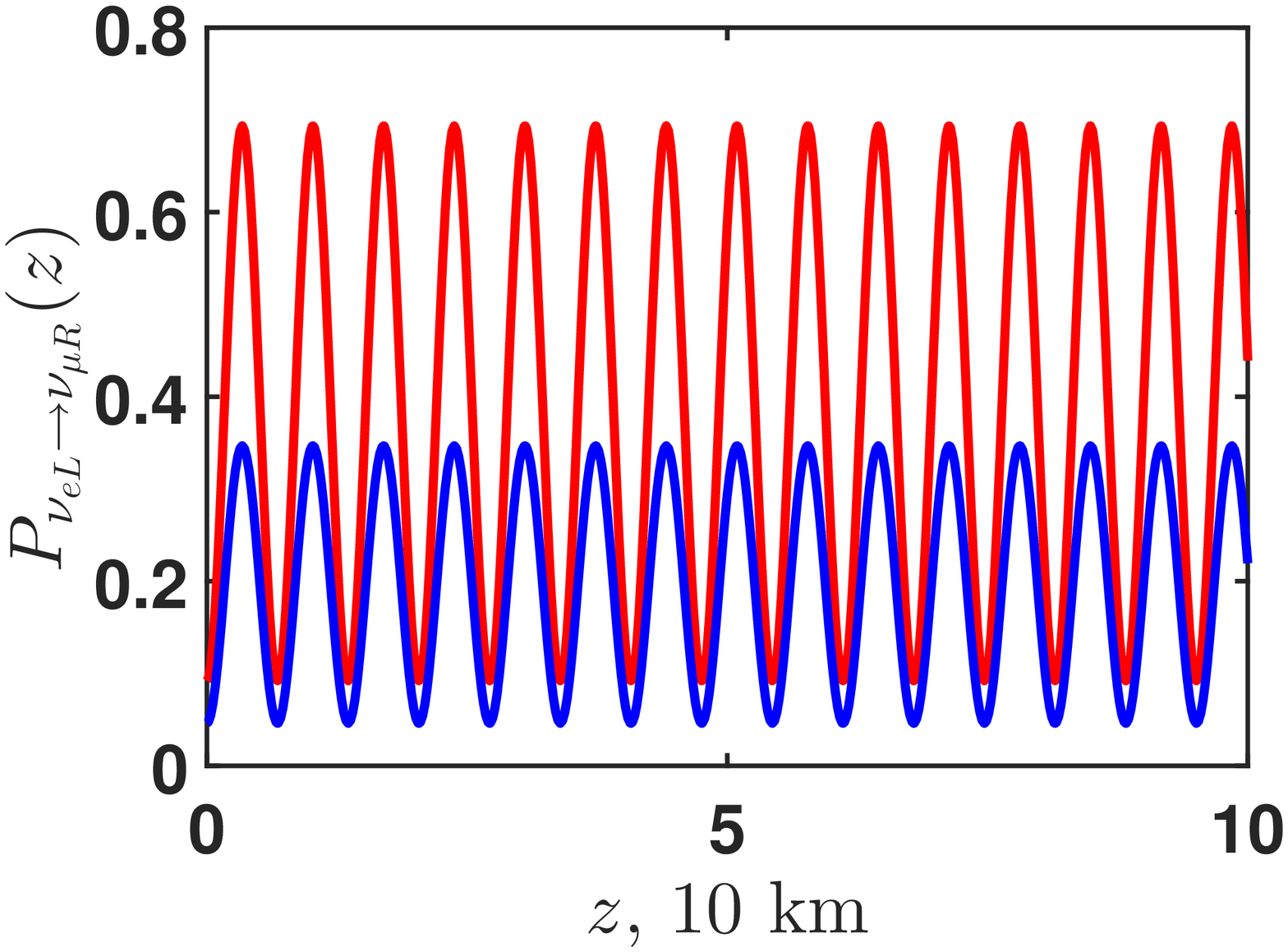}}
  \hskip-.6cm
  \subfigure[]
  {\label{2b}
  \includegraphics[scale=.35]{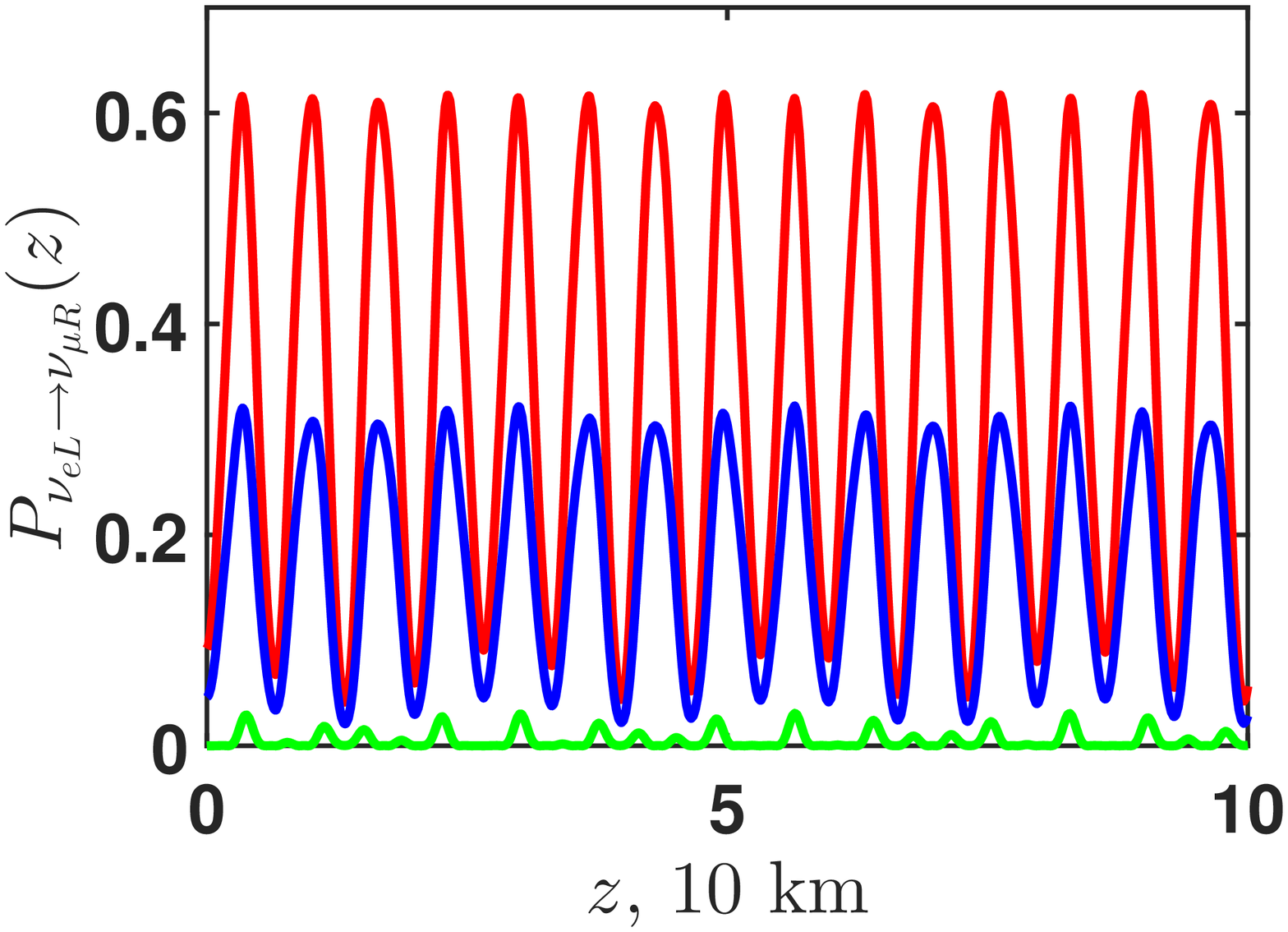}}
  \\
  \subfigure[]
  {\label{2c}
  \includegraphics[scale=.35]{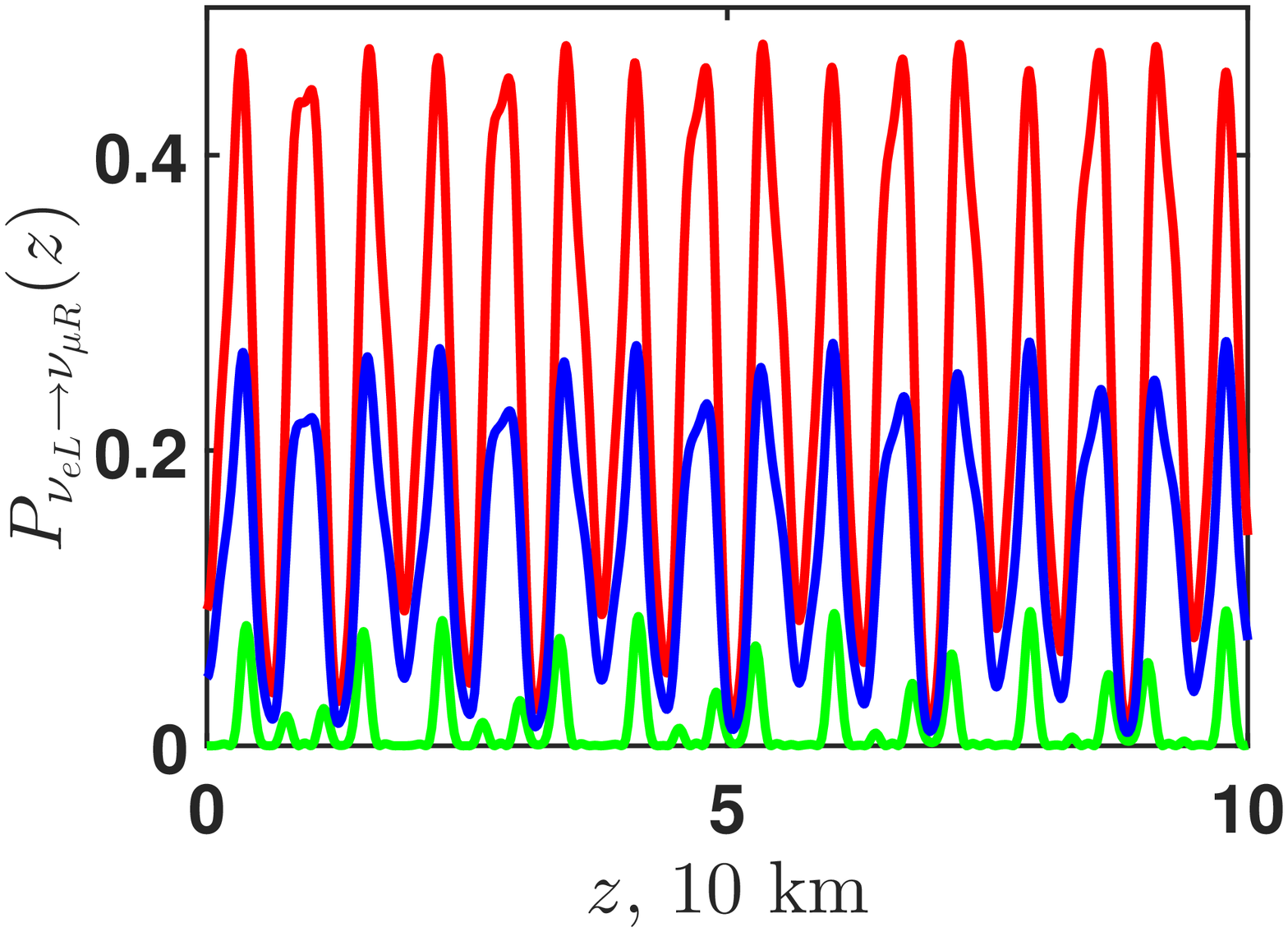}}
  \hskip-.6cm
  \subfigure[]
  {\label{2d}
  \includegraphics[scale=.35]{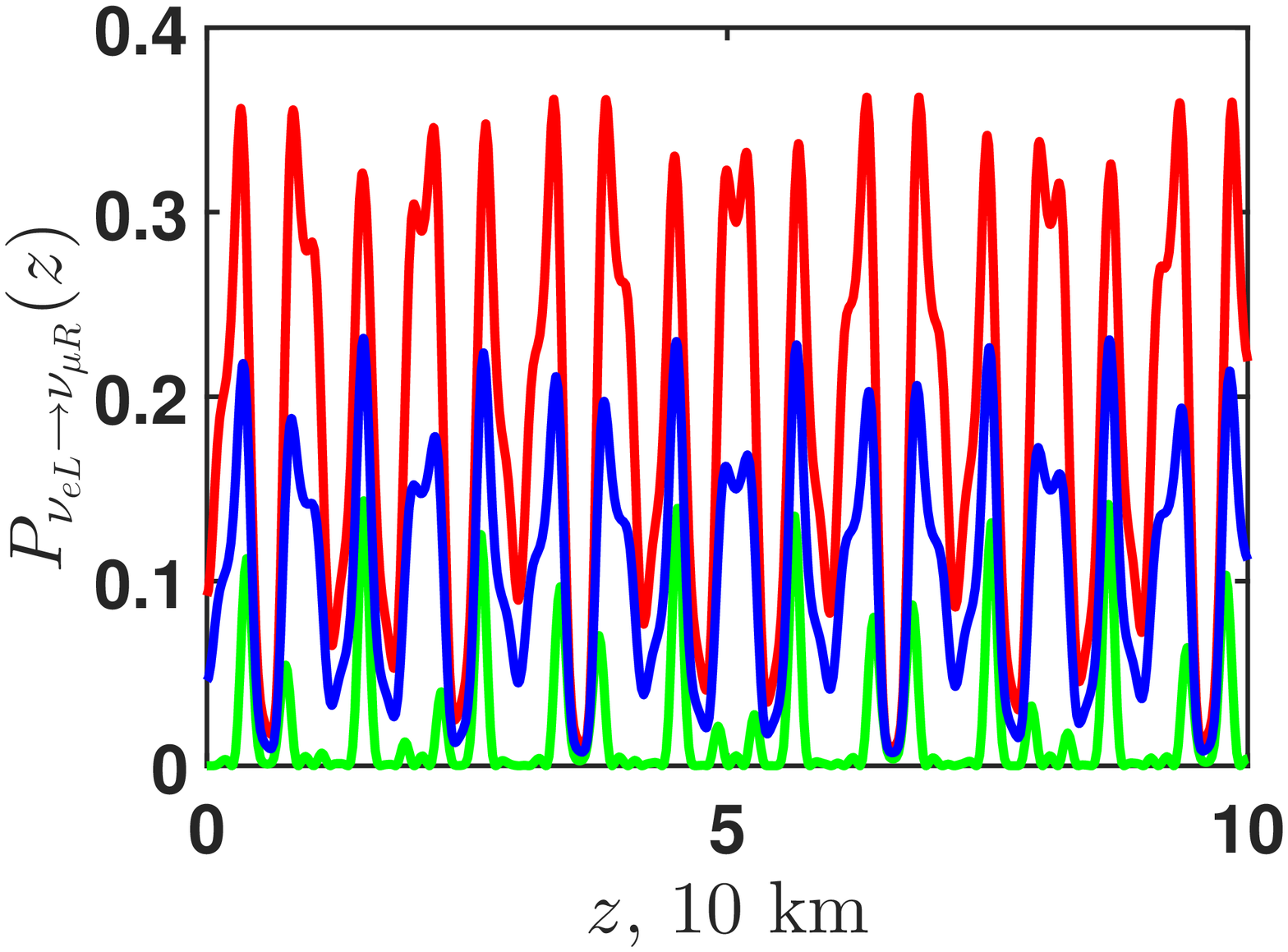}}
  \protect
  \caption{The upper (red lines) and lower (green lines) envelope functions, as well as the averaged 
  transition probability (blue lines), corresponding to Eq.~(\ref{eq:PLRtildea}), for
  $\nu_{e\mathrm{L}}\to\nu_{\mu\mathrm{R}}$ oscillations in the electromagnetic
  wave versus the distance $z$ passed by the neutrino beam for different
  diagonal magnetic moment $\mu'$, based on the numerical solution
  of Eq.~(\ref{eq:SchPsitildeDMM}). The parameters of the neutrino system
  and $B_{0}$ are the same as in Fig.~\ref{fig:PTMM}. The frequency
  of the electromagnetic wave is $\omega=10^{12}\,\text{s}^{-1}$.
  (a)~$\mu'=0$ ($\epsilon_{0}=0$);
  (b)~$\mu'=2.8\times10^{-3}\mu=2.8\times10^{-14}\mu_\mathrm{B}$
  ($\epsilon_{0}=0.1$);
  (c)~$\mu'=5.5\times10^{-3}\mu=5.5\times10^{-14}\mu_\mathrm{B}$
  ($\epsilon_{0}=0.2$);
  (d)~$\mu'=8.3\times10^{-3}\mu=8.3\times10^{-14}\mu_\mathrm{B}$
  ($\epsilon_{0}=0.3$).\label{fig:PDMM12}}
\end{figure}

\begin{figure}
  \centering
  \subfigure[]
  {\label{3a}
  \includegraphics[scale=.35]{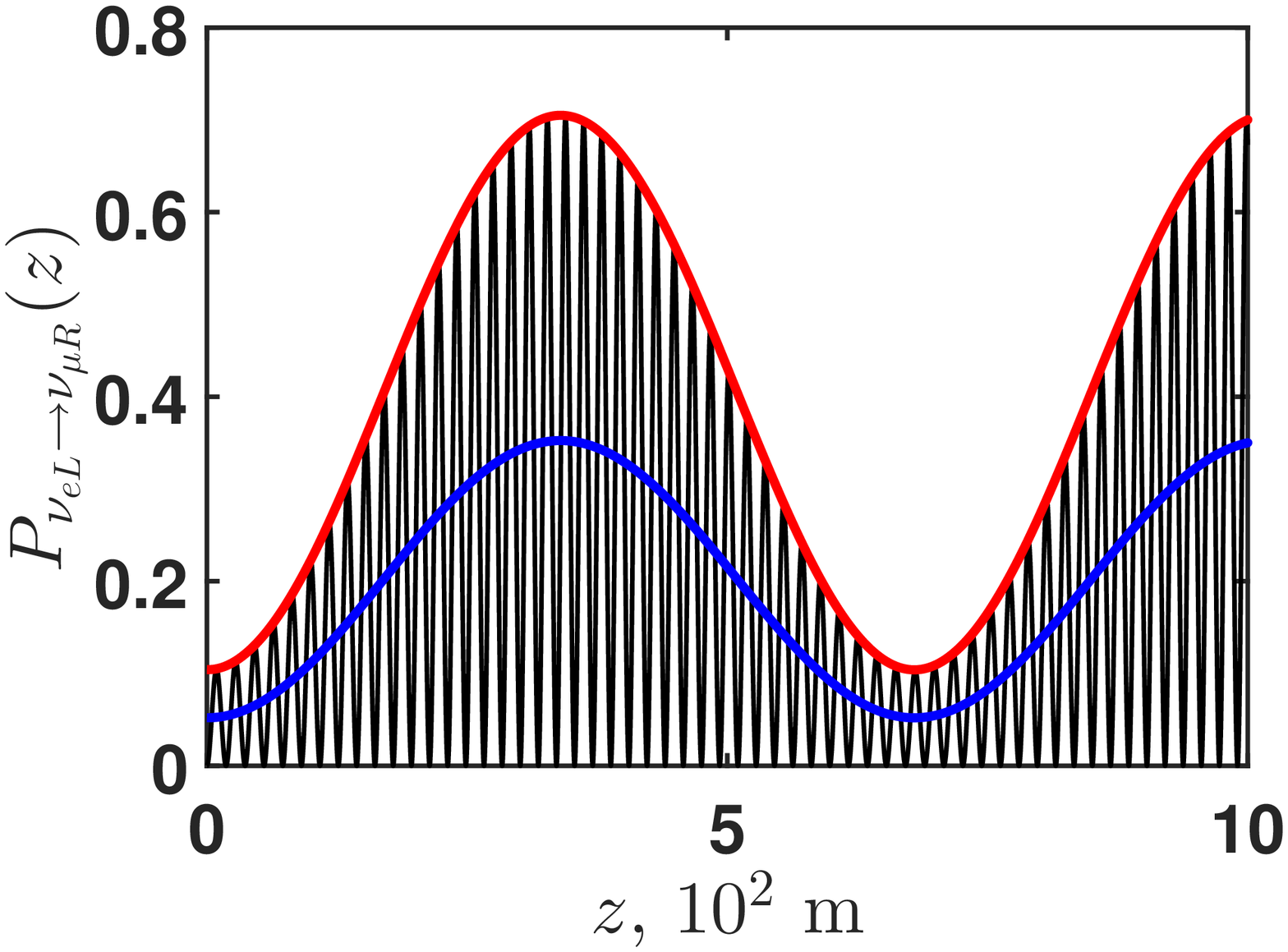}}
  \hskip-.6cm
  \subfigure[]
  {\label{3b}
  \includegraphics[scale=.35]{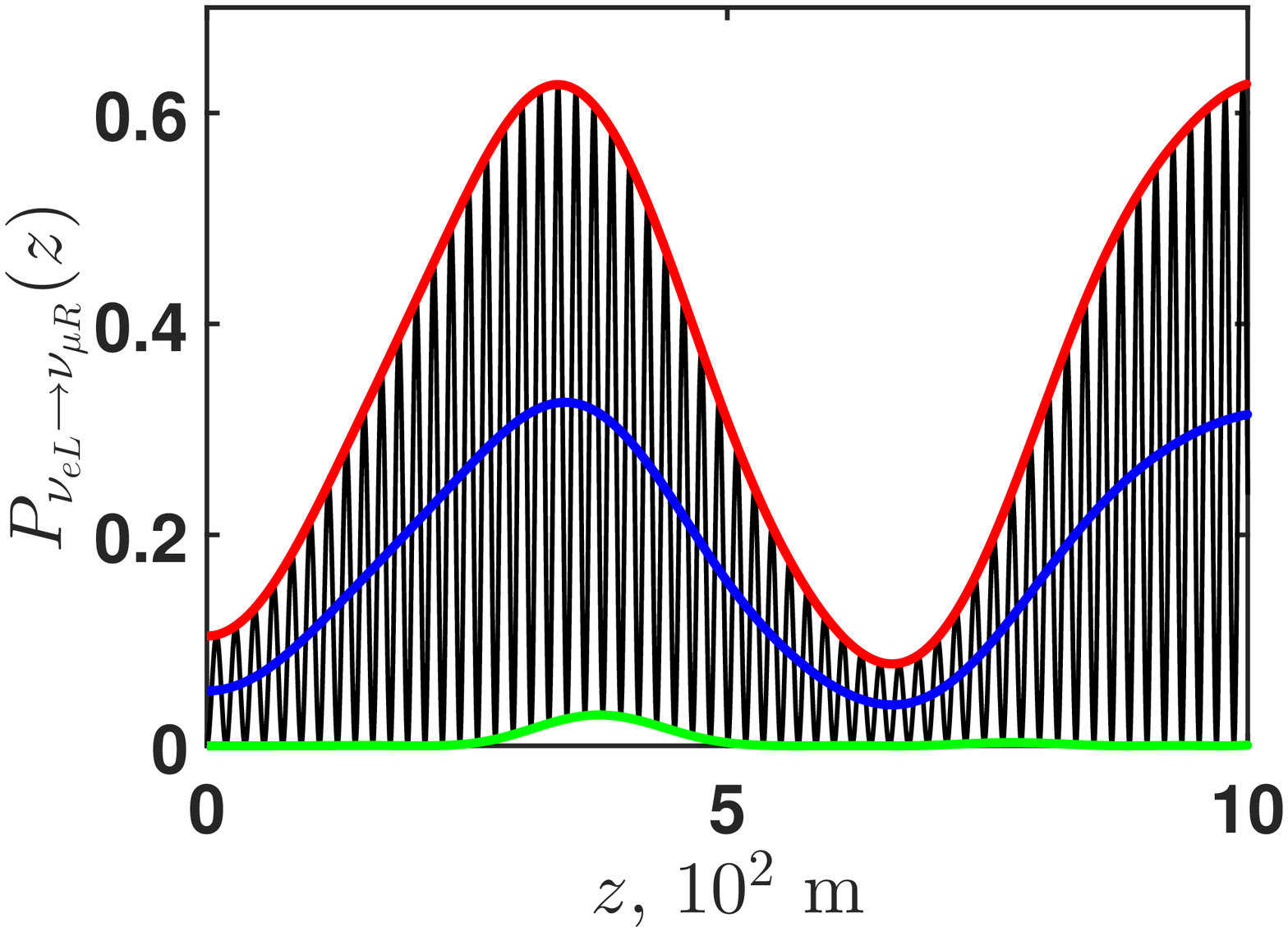}}
  \\
  \subfigure[]
  {\label{3c}
  \includegraphics[scale=.35]{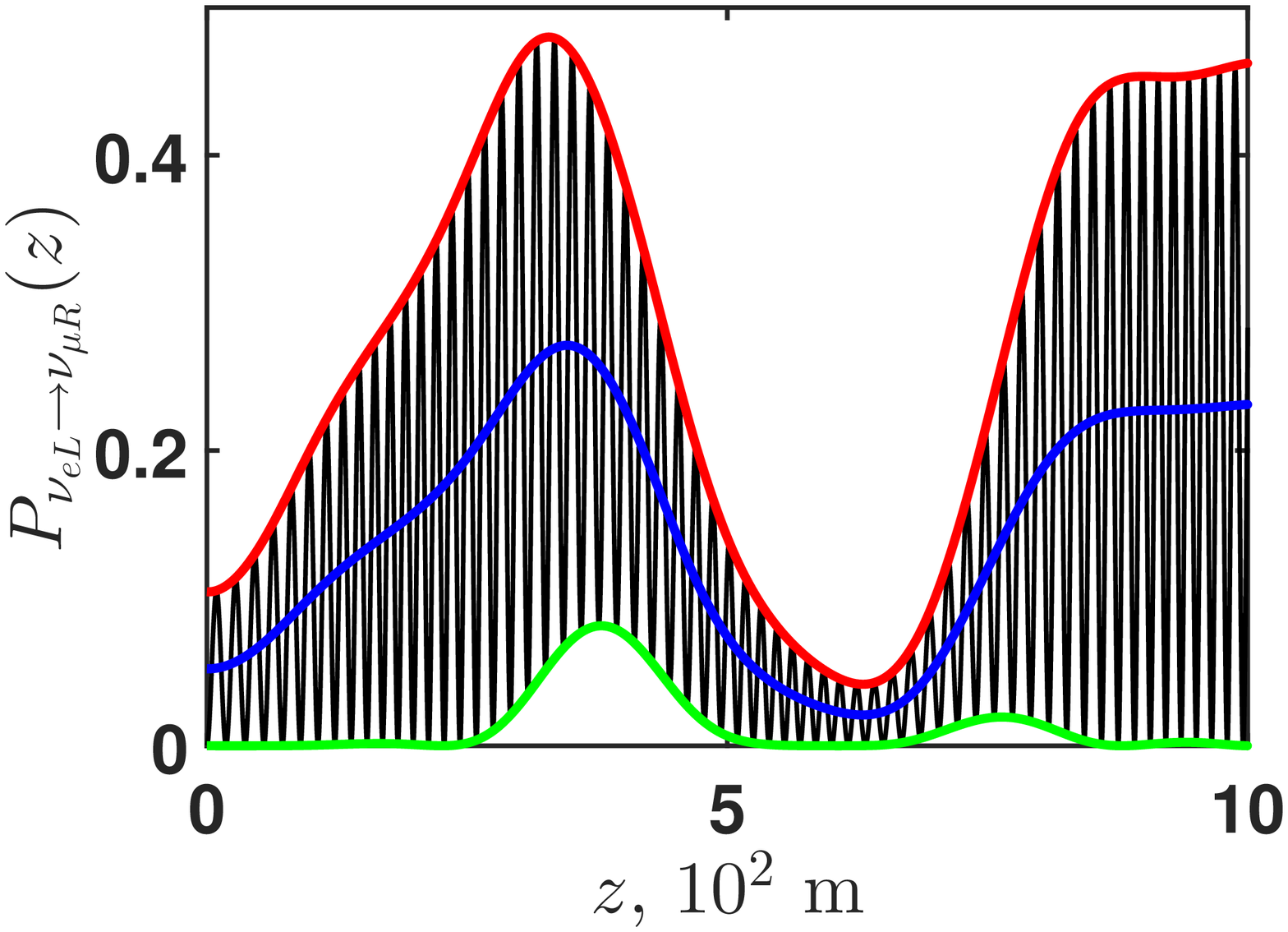}}
  \hskip-.6cm
  \subfigure[]
  {\label{3d}
  \includegraphics[scale=.35]{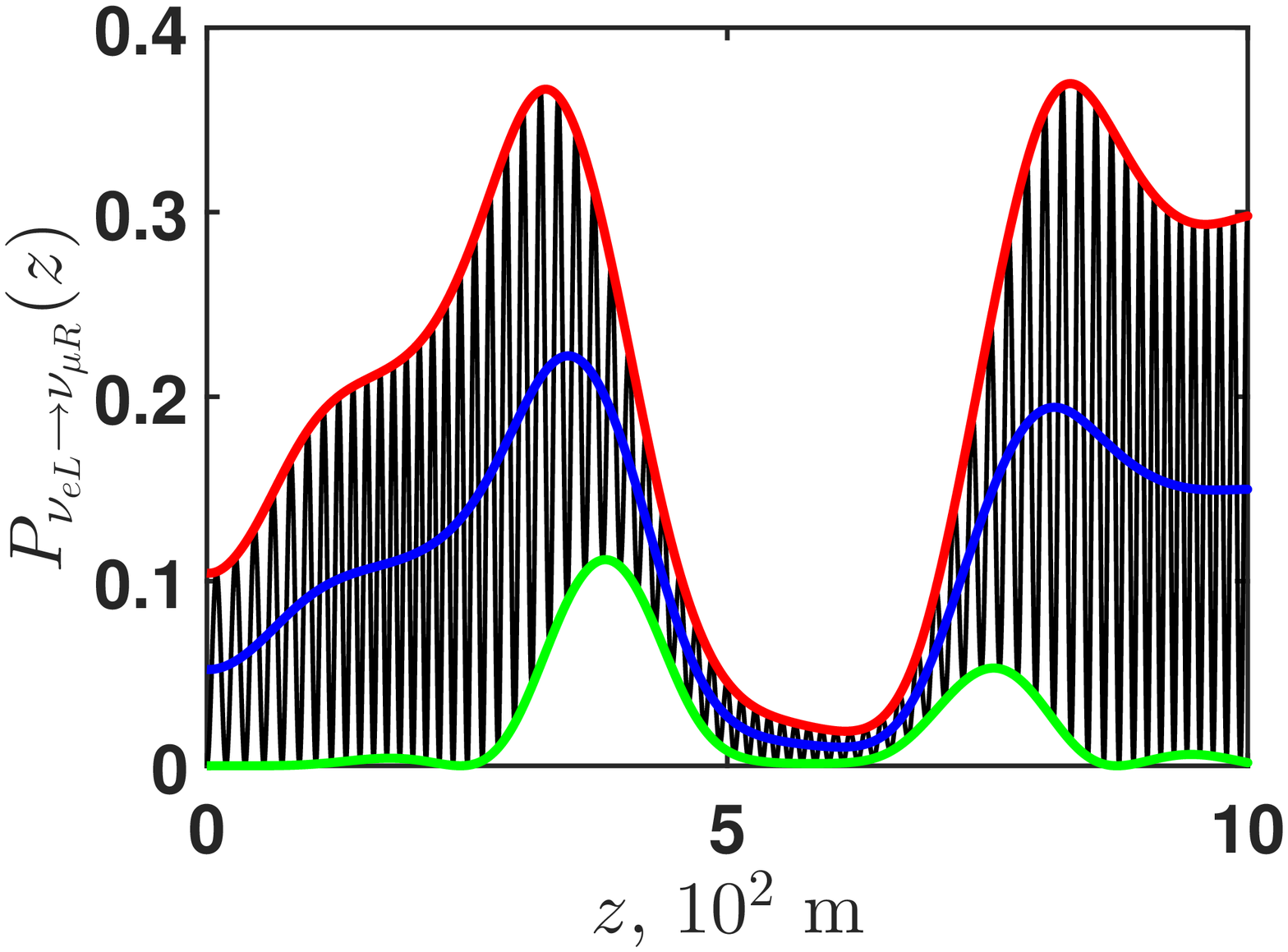}}
  \protect
  \caption{The transition probability (black lines) for
  $\nu_{e\mathrm{L}}\to\nu_{\mu\mathrm{R}}$
  oscillations, as well as the upper (red lines) and lower (green lines) envelope functions,
  and the averaged 
  transition probability (blue lines) for this oscillations channel, based on the numerical solution of
  Eq.~\eqref{eq:SchPsitildeDMM}. The parameters of the neutrino system and the 
  electromagnetic wave are the same as in Fig.~\ref{fig:PDMM12}, except for the frequency,
  which is now equal to $\omega=10^{13}\,\text{s}^{-1}$.  
  \label{fig:PDMM13}}
\end{figure}

For the illustration of the influence of the diagonal magnetic moments
on neutrino spin-flavor oscillations in a plane electromagnetic wave,
we consider $\nu_{e\mathrm{L}}\to\nu_{\mu\mathrm{R}}$ with the same
parameters as in Sec.~\ref{sec:TMM}.
In Figs.~\ref{fig:PDMM12} and~\ref{fig:PDMM13},
we show $P_{\nu_{e\mathrm{L}}\to\nu_{\mu\mathrm{R}}}$ based on the
numerical solution of Eq.~(\ref{eq:SchPsitildeDMM}) for different
$\mu'$ (or $\epsilon_{0}=\mu'B_{0}/\omega$) and $\omega$.
The upper and lower envelope functions, as well as the averaged transition
probability, are also depicted in Figs.~\ref{fig:PDMM12} and~\ref{fig:PDMM13}. To build these envelope functions we use the spline interpolation of the
maxima and minima of $P_{\nu_{e\mathrm{L}}\to\nu_{\mu\mathrm{R}}}$
respectively.

Note that, as in Fig.~\ref{1a}, we do not represent the evolution of $P_{\nu_{e\mathrm{L}}\to\nu_{\mu\mathrm{R}}}(z)$ in Fig.~\ref{fig:PDMM12} since this function is rapidly oscillating and, hence, is indistinguishable. Instead, we just show the envelope functions and the averaged transition probability in Fig.~\ref{fig:PDMM12}.

One can see that $P_{\nu_{e\mathrm{L}}\to\nu_{\mu\mathrm{R}}}$ and the envelope functions in
Figs.~\ref{2a} and~\ref{3a}, corresponding to $\mu'=0$, coincide with the curves shown in Figs.~\ref{1a} and~\ref{1b},
which were built on the basis of the exact solution of Eq.~(\ref{eq:SchPsitilde});
cf. Eqs.~(\ref{eq:tildePsisol}), (\ref{eq:UV}) and~(\ref{eq:PbetaLalphaR}).
It is also interesting to note that $P_{\nu_{e\mathrm{L}}\to\nu_{\mu\mathrm{R}}}$ and the envelope functions 
for spin-flavor oscillations of neutrinos with $\mu'\sim10^{-2}\mu\ll\mu$,
shown in Figs.~\ref{2d} and~\ref{3d}, significantly differs from the case $\mu'=0$ shown in Figs.~\ref{2a} and~\ref{3a}.

Basing on Figs.~\ref{2b}-\ref{2d} and~\ref{3b}-\ref{3d}, one can conclude that the averaged transition
probability diminishes if one accounts for $\mu'>0$ in the system.
Thus the situation when $\mu_{a} = 0$, studied in Sec.~\ref{sec:TMM},
is more preferable from the point of view of phenomenological applications.

\begin{figure}
  \centering
  \subfigure[]
  {\label{5a}
  \includegraphics[scale=.35]{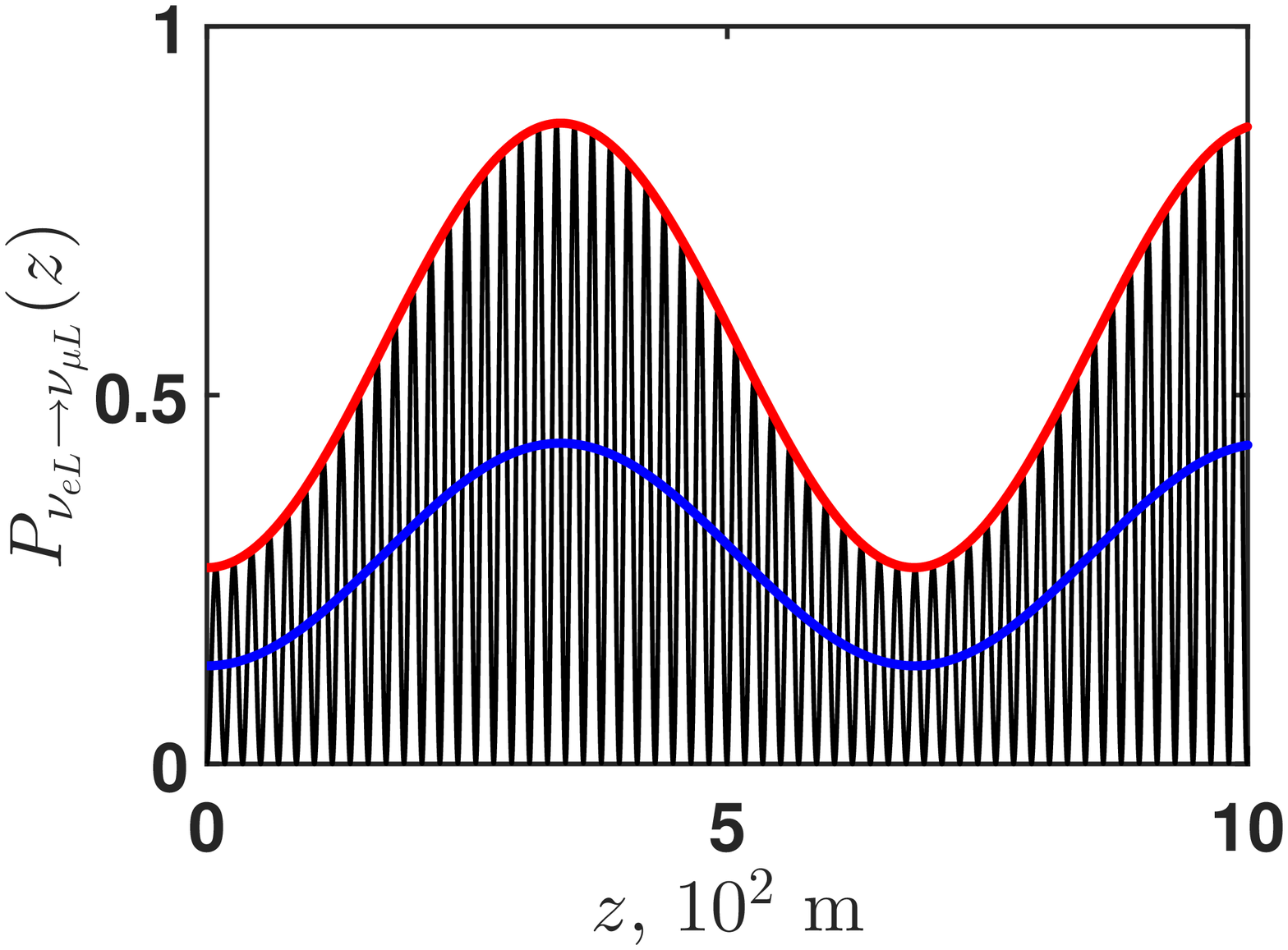}}
  \hskip-.6cm
  \subfigure[]
  {\label{5b}
  \includegraphics[scale=.35]{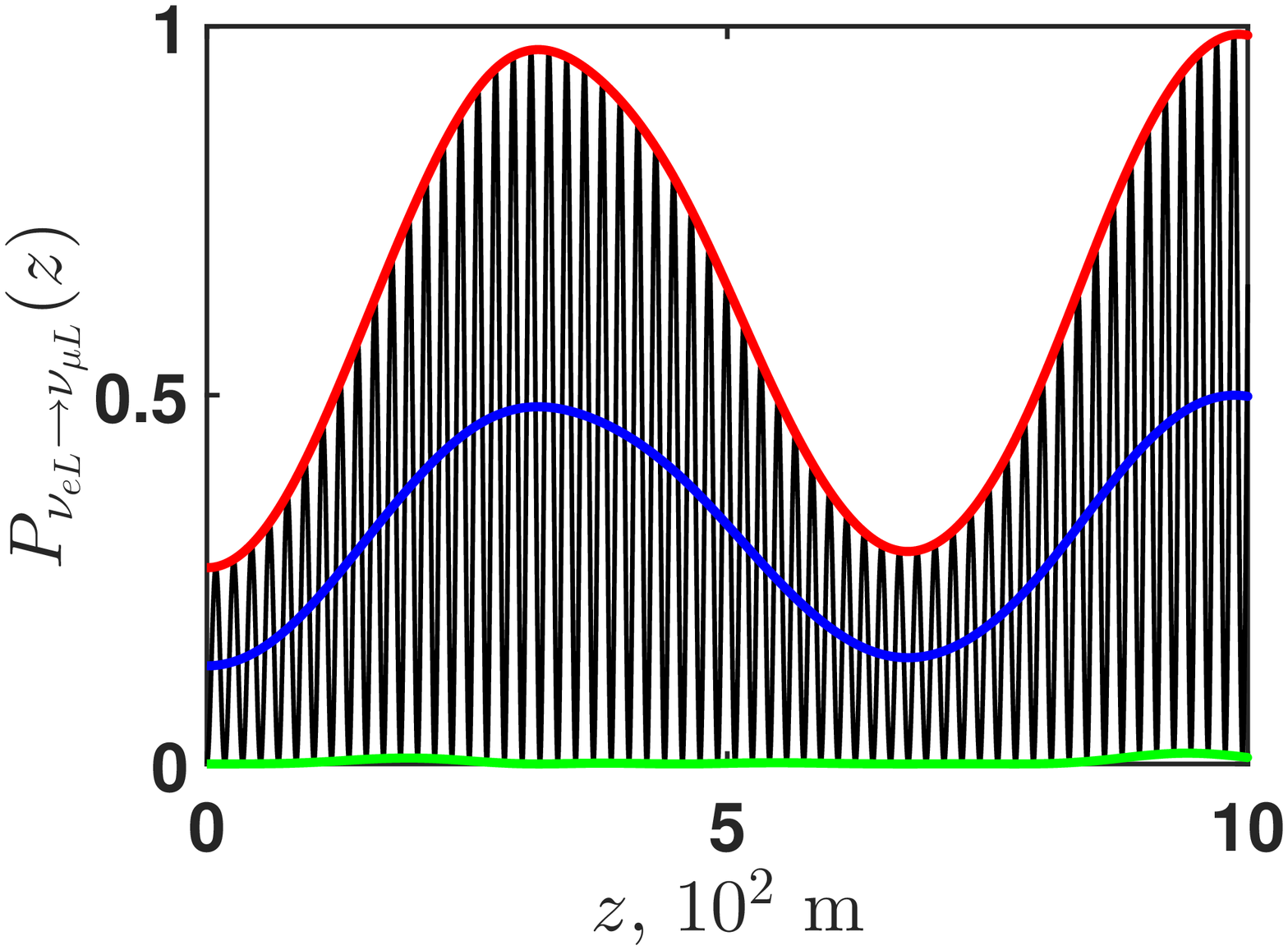}}
  \\
  \subfigure[]
  {\label{5c}
  \includegraphics[scale=.35]{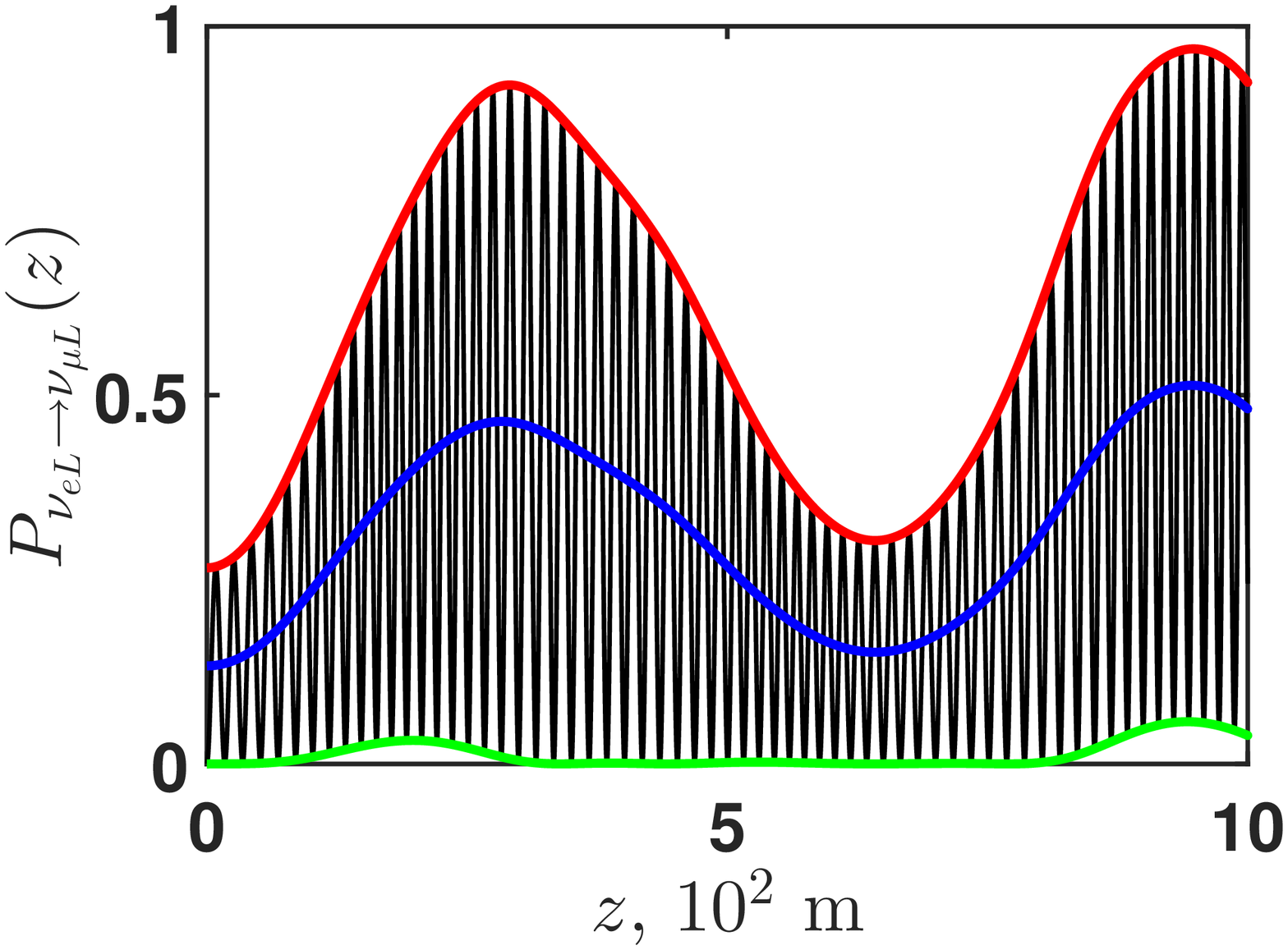}}
  \hskip-.6cm
  \subfigure[]
  {\label{5d}
  \includegraphics[scale=.35]{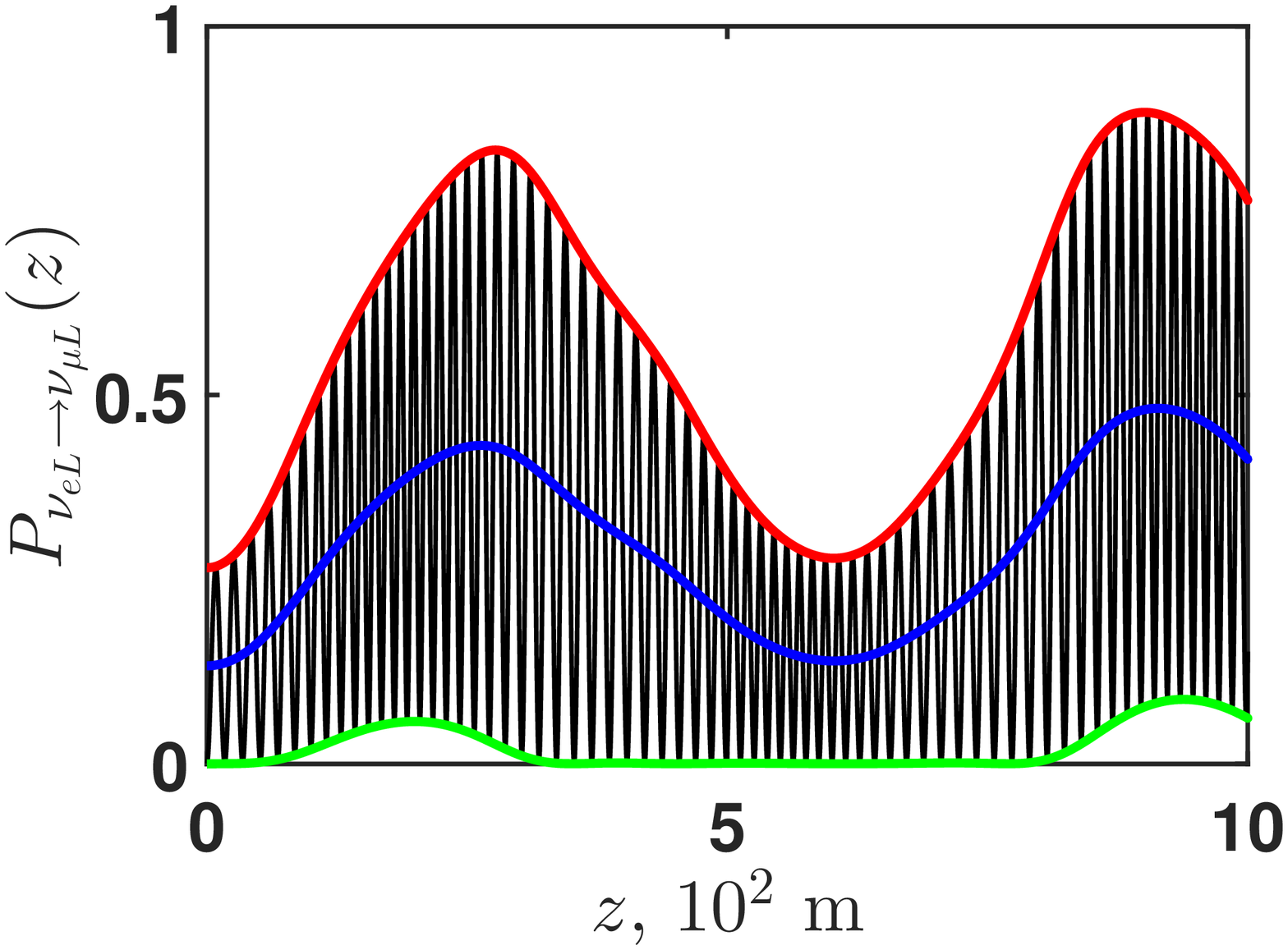}}
  \protect
  \caption{The transition probability (black lines) for
  $\nu_{e\mathrm{L}}\to\nu_{\mu\mathrm{L}}$
  oscillations, as well as the upper (red lines) and lower (green lines) envelope functions,
  and the averaged 
  transition probability (blue lines) for this oscillations channel, based on the numerical solution of
  Eq.~\eqref{eq:SchPsitildeDMM}. The parameters of the neutrino system and the 
  electromagnetic wave are the same as in Fig.~\ref{fig:PDMM13}.  
  \label{fig:eLmuL13}}
\end{figure}

\begin{figure}
  \centering
  \subfigure[]
  {\label{6a}
  \includegraphics[scale=.35]{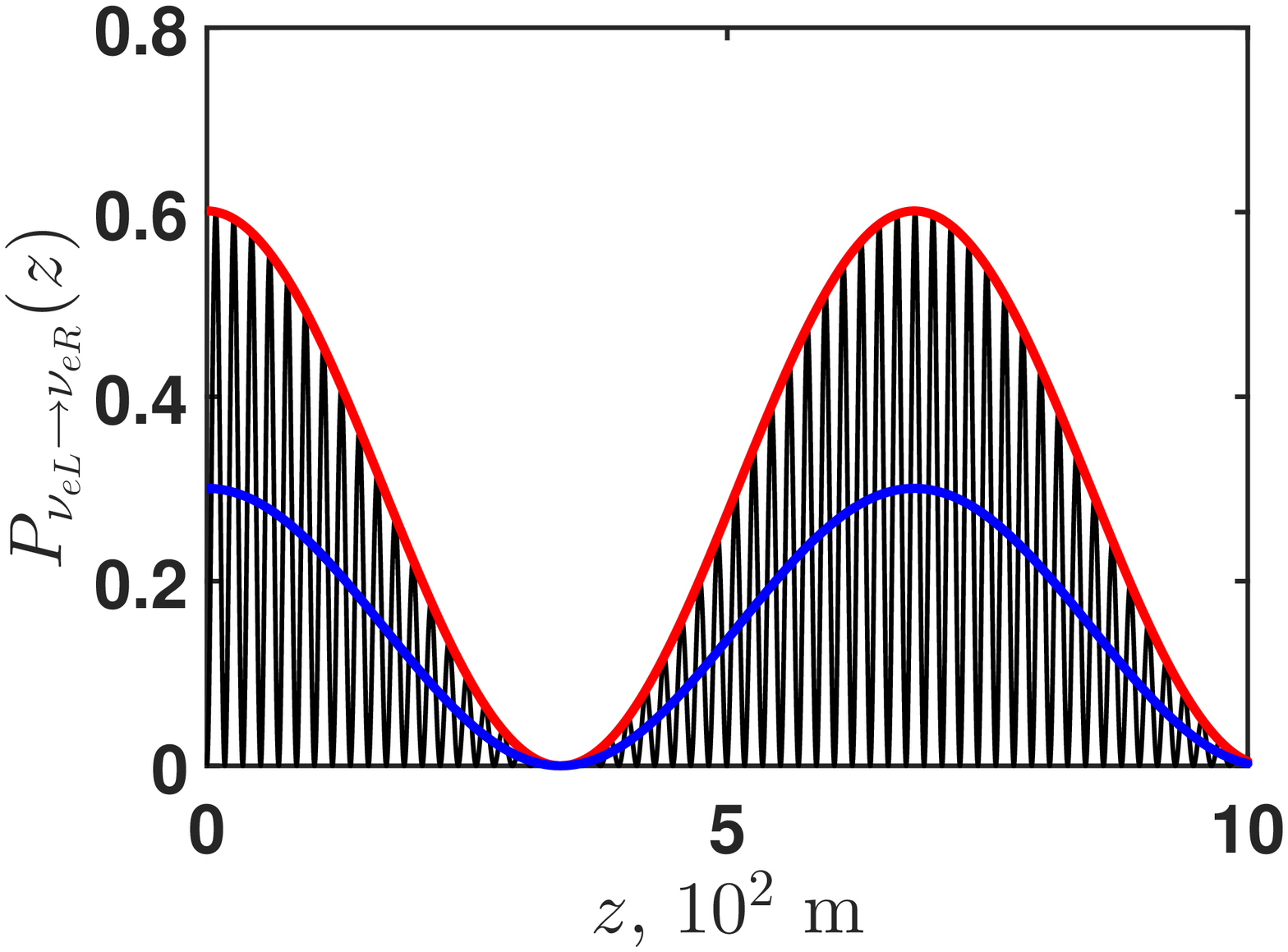}}
  \hskip-.6cm
  \subfigure[]
  {\label{6b}
  \includegraphics[scale=.35]{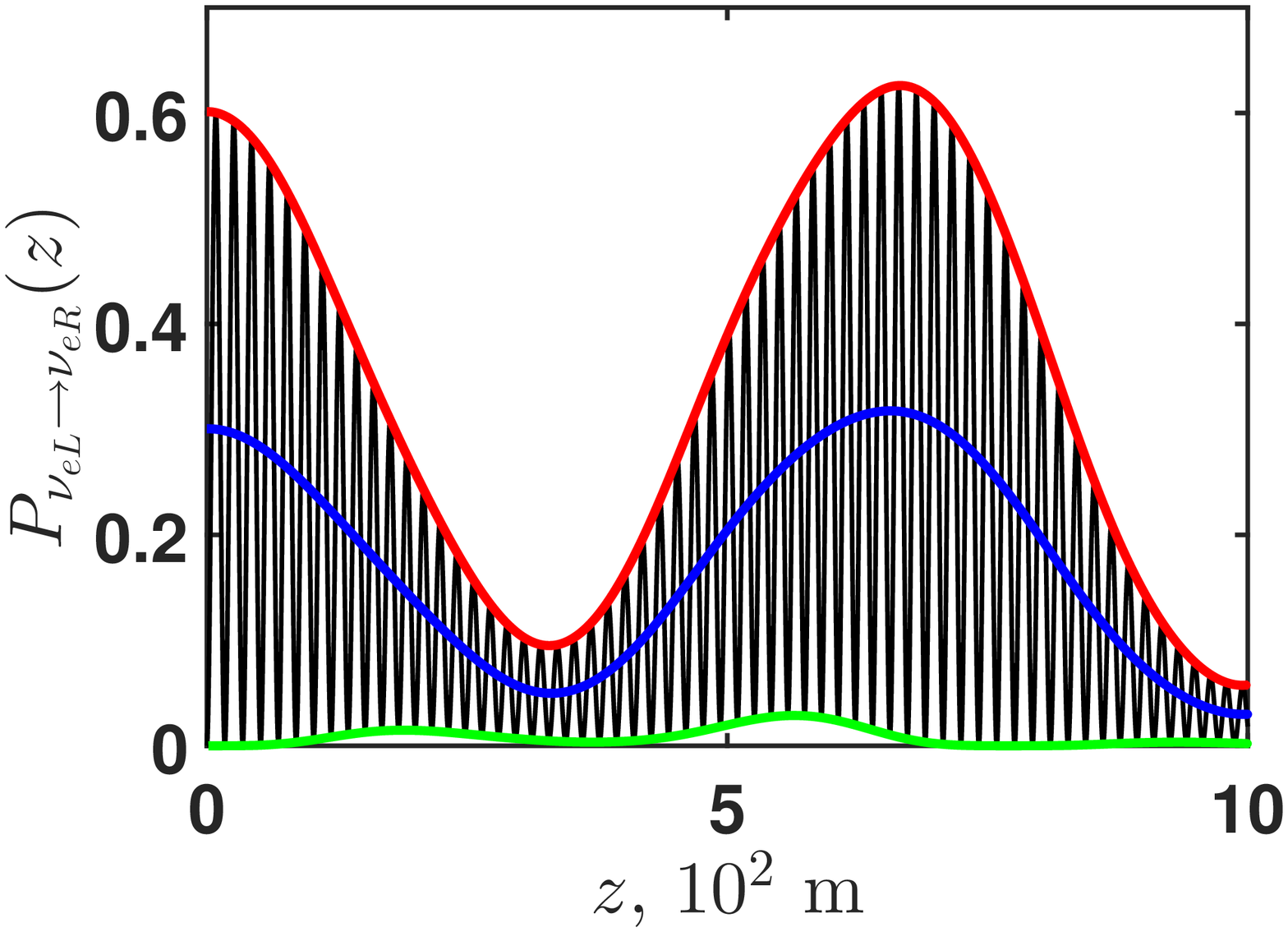}}
  \\
  \subfigure[]
  {\label{6c}
  \includegraphics[scale=.35]{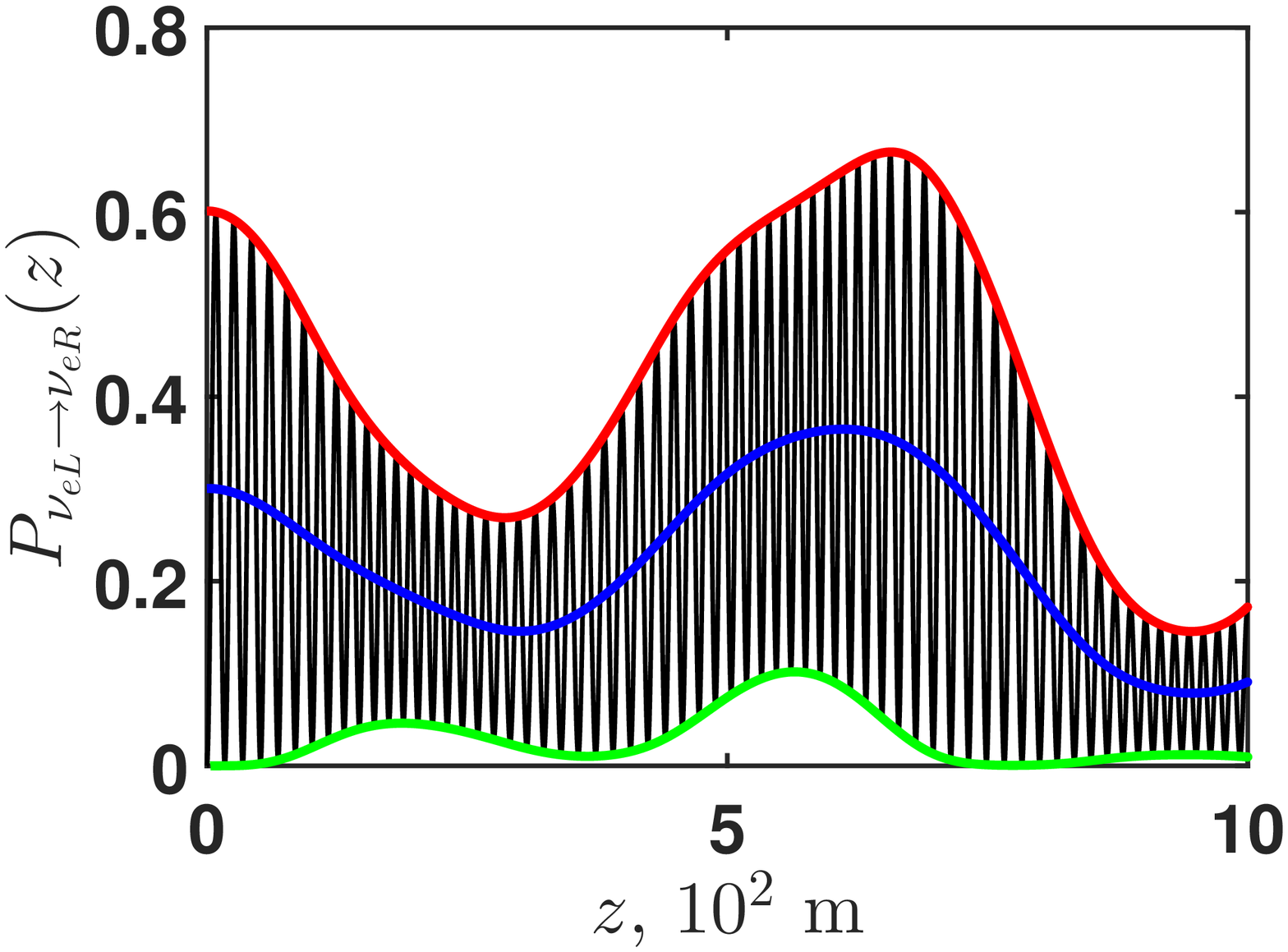}}
  \hskip-.6cm
  \subfigure[]
  {\label{6d}
  \includegraphics[scale=.35]{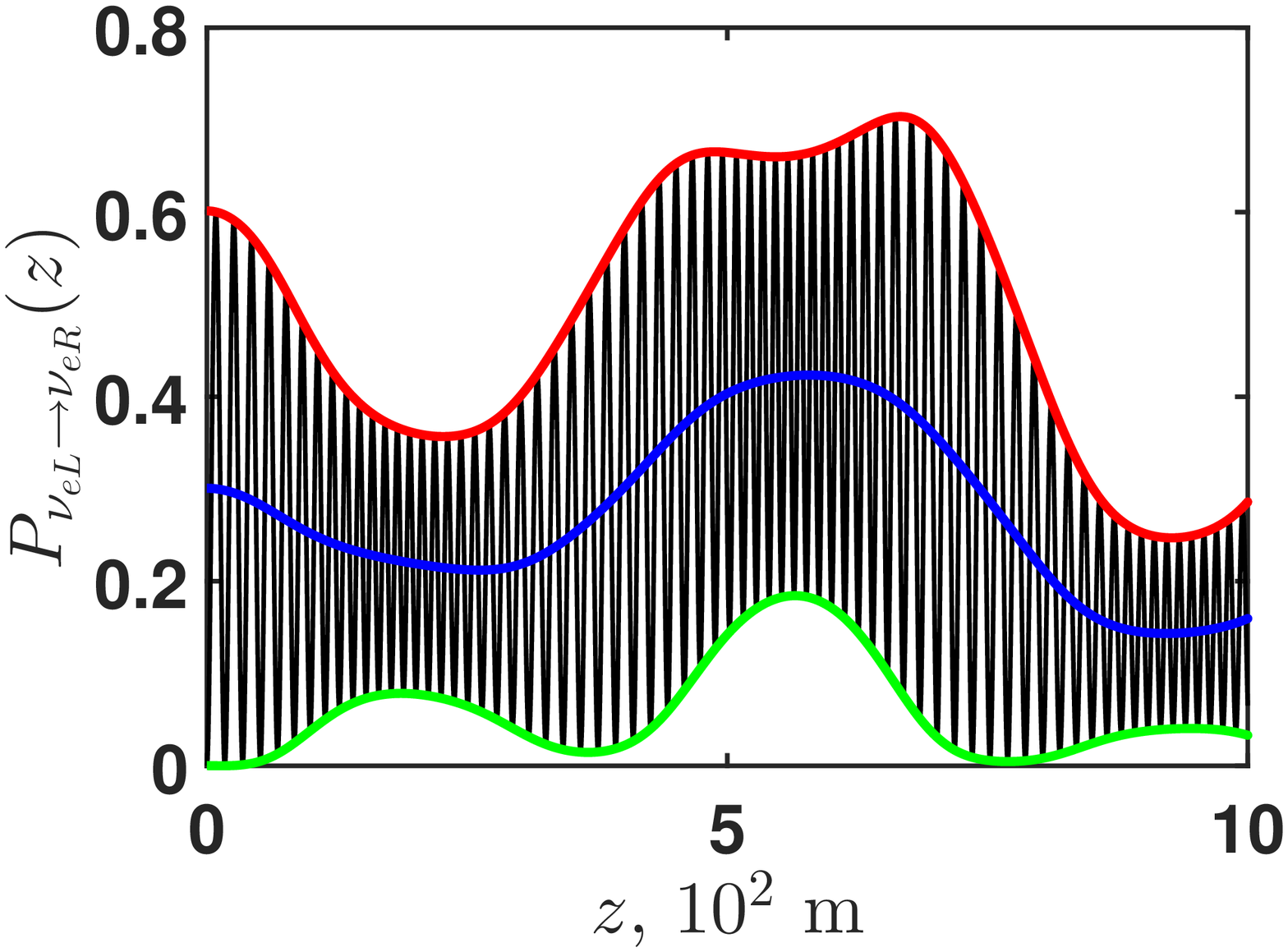}}
  \protect
  \caption{The transition probability (black lines) for
  $\nu_{e\mathrm{L}}\to\nu_{e\mathrm{R}}$
  oscillations, as well as the upper (red lines) and lower (green lines) envelope functions,
  and the averaged 
  transition probability (blue lines) for this oscillations channel, based on the numerical solution of
  Eq.~\eqref{eq:SchPsitildeDMM}. The parameters of the neutrino system and the 
  electromagnetic wave are the same as in Fig.~\ref{fig:PDMM13}.  
  \label{fig:eLeR13}}
\end{figure}

\begin{figure}
  \centering
  \subfigure[]
  {\label{7a}
  \includegraphics[scale=.35]{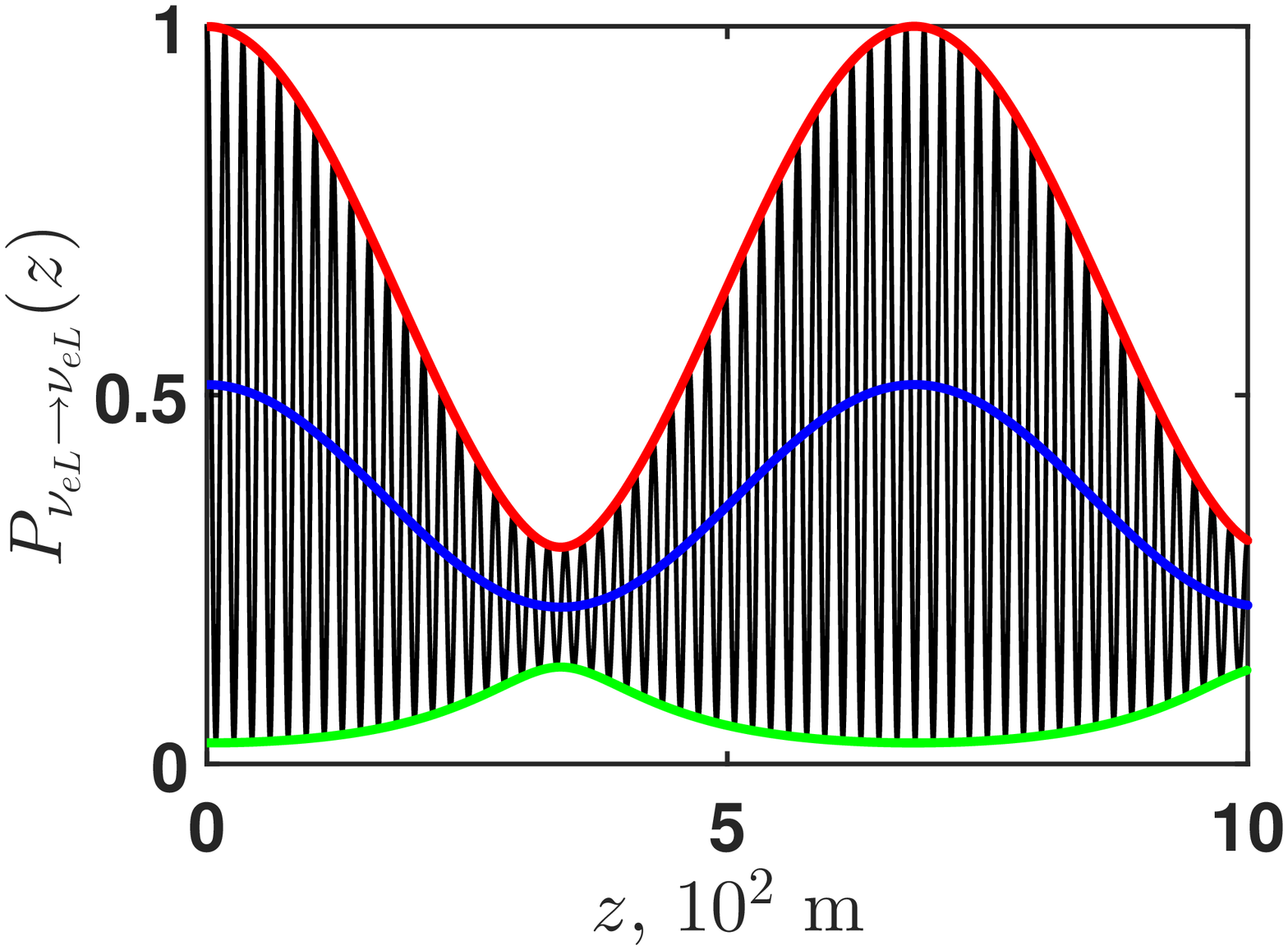}}
  \hskip-.6cm
  \subfigure[]
  {\label{7b}
  \includegraphics[scale=.35]{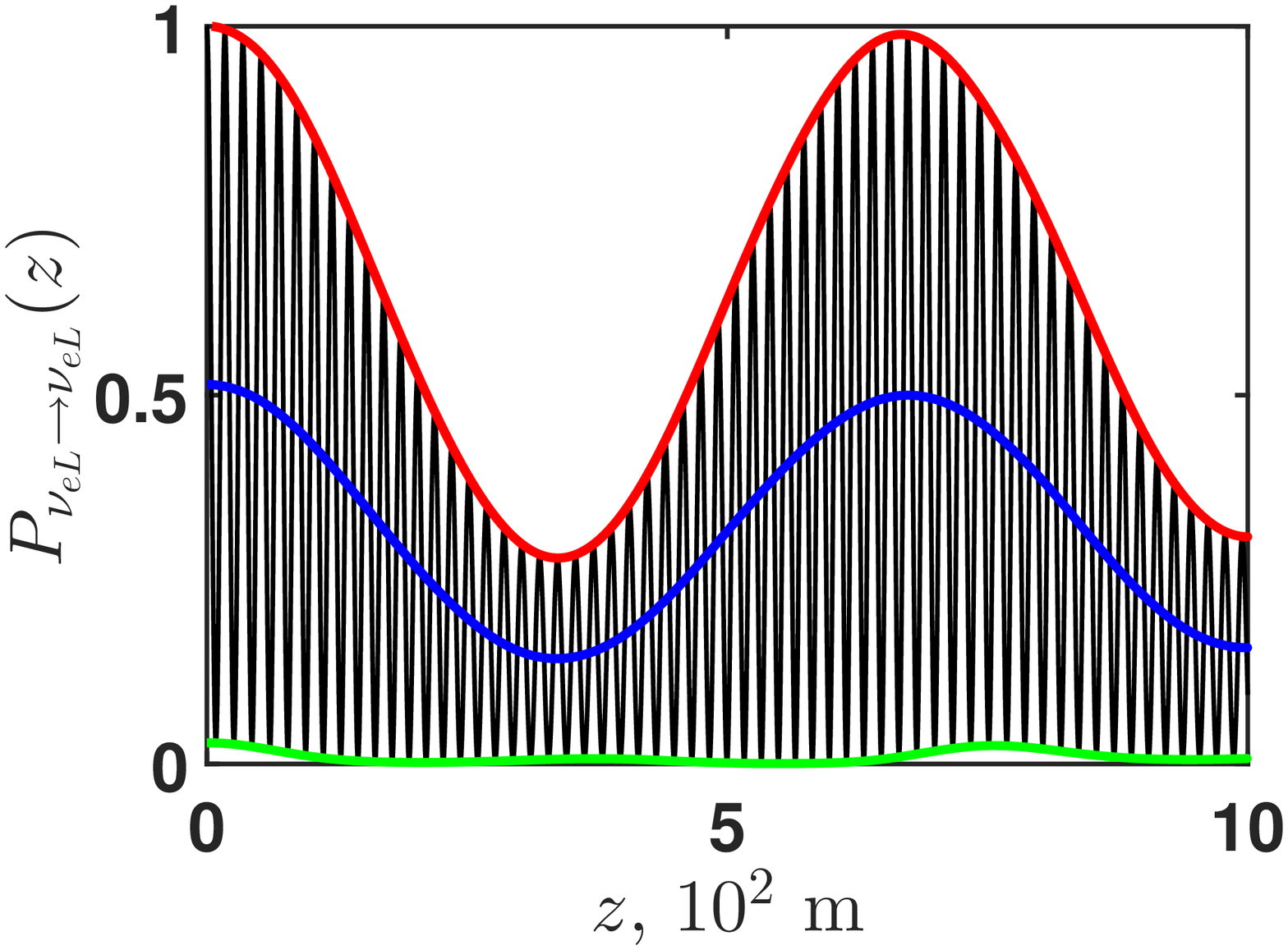}}
  \\
  \subfigure[]
  {\label{7c}
  \includegraphics[scale=.35]{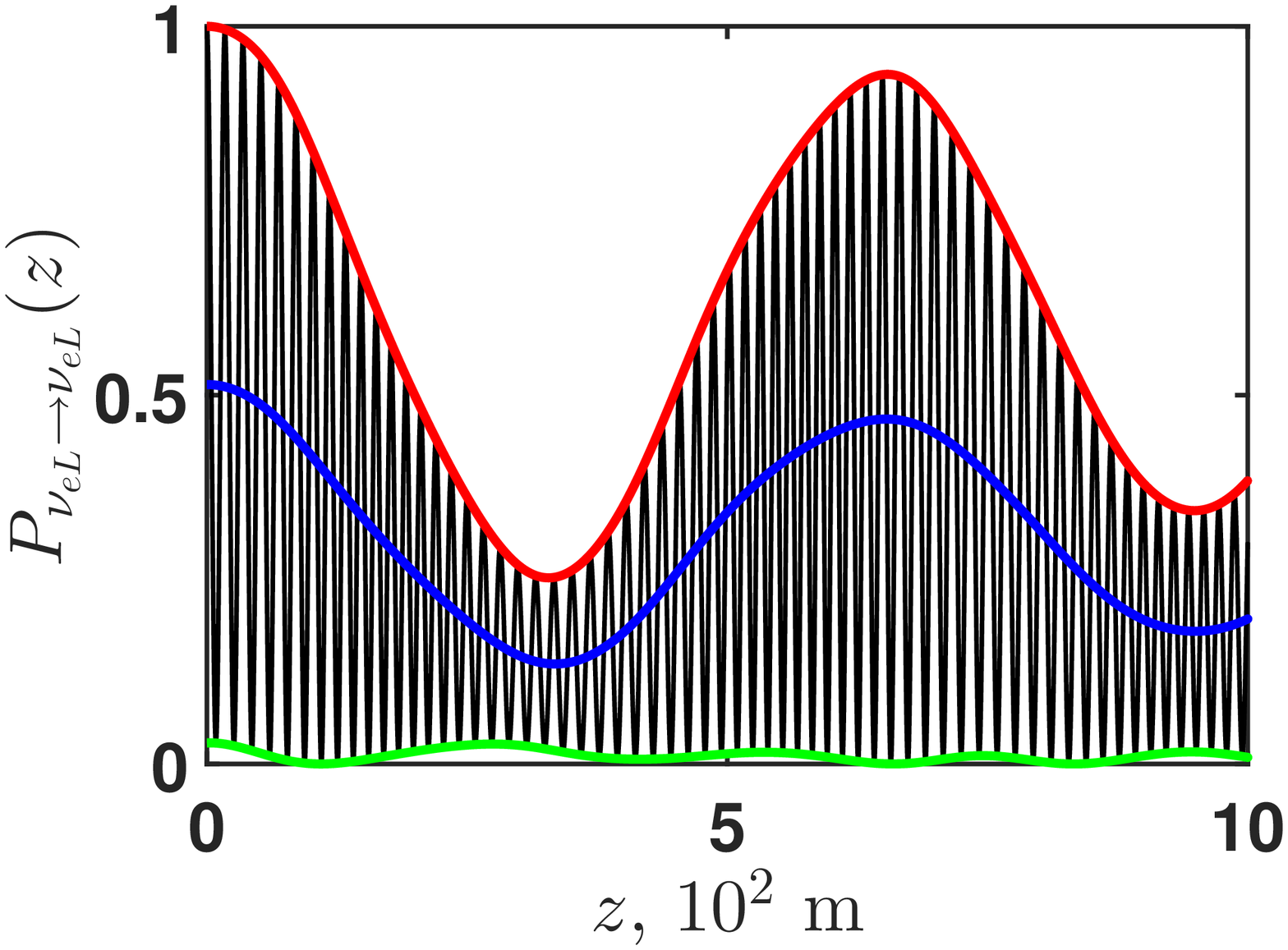}}
  \hskip-.6cm
  \subfigure[]
  {\label{7d}
  \includegraphics[scale=.35]{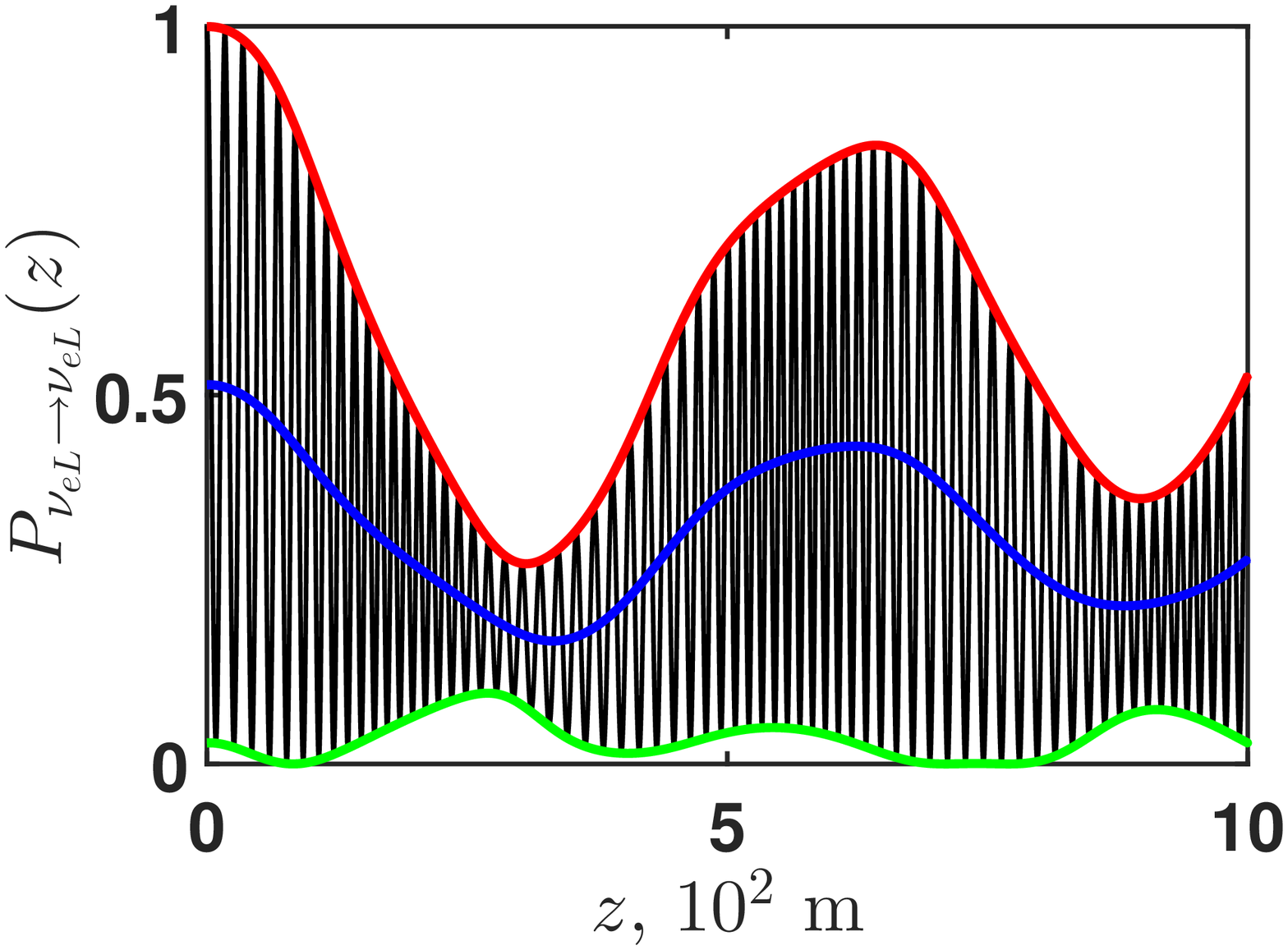}}
  \protect
  \caption{The survival probability (black lines) for
  $\nu_{e\mathrm{L}}\to\nu_{e\mathrm{L}}$
  oscillations, as well as the upper (red lines) and lower (green lines) envelope functions,
  and the averaged 
  survival probability (blue lines) for this oscillations channel, based on the numerical solution of
  Eq.~\eqref{eq:SchPsitildeDMM}. The parameters of the neutrino system and the 
  electromagnetic wave are the same as in Fig.~\ref{fig:PDMM13}.  
  \label{fig:eLeL13}}
\end{figure}

Besides $\nu_{e\mathrm{L}}\to\nu_{\mu\mathrm{R}}$ oscillations, we can study other channels of neutrino oscillations on the basis of the numerical solution of Eq.~\eqref{eq:SchPsitildeDMM}. For example, in Fig.~\ref{fig:eLmuL13}, we demonstrate the transition probability, the upper and lower envelope functions, and the averaged transition probability for $\nu_{e\mathrm{L}}\to\nu_{\mu\mathrm{L}}$ oscillations. These dependencies for $\nu_{e\mathrm{L}}\to\nu_{e\mathrm{R}}$ oscillations are shown in Fig.~\ref{fig:eLeR13}. The survival probability for oscillations $\nu_{e\mathrm{L}}\to\nu_{e\mathrm{L}}$, the upper and lower envelope functions, and the averaged survival probability are given in Fig.~\ref{fig:eLeL13}.

One can see in Figs.~\ref{fig:PDMM13}-\ref{fig:eLeR13} that $P_{\nu_{e\mathrm{L}}\to\nu_{\mu\mathrm{R}}} = P_{\nu_{e\mathrm{L}}\to\nu_{\mu\mathrm{L}}} = P_{\nu_{e\mathrm{L}}\to\nu_{e\mathrm{R}}} = 0$ at $z=0$, whereas $P_{\nu_{e\mathrm{L}}\to\nu_{e\mathrm{L}}}(z=0) = 1$ in Fig.~\ref{fig:eLeL13}, i.e. the initial condition is fulfilled. Figs.~\ref{fig:PDMM13}-\ref{fig:eLeL13} are built for $\omega=10^{13}\,\text{s}^{-1}$. The analogues of these dependencies for $\omega=10^{12}\,\text{s}^{-1}$ are not presented here since the transition and survival probabilities are rapidly oscillating functions and, hence, are quite difficult for the analysis.

We also mention that, in this section, we study only the situation when $\omega \geq 10^{12}\,\text{s}^{-1}$.
The numerical solution of Eq.~(\ref{eq:SchPsitildeDMM}) for $\omega < 10^{12}\,\text{s}^{-1}$ (i.e. the situation corresponding to electromagnetic waves emitted by a radio pulsar; cf. Sec.~\ref{sec:TMM}) is quite problematic
since Eq.~(\ref{eq:SchPsitildeDMM}) becomes extremely stiff then.

\section{Spin-flavor oscillations: great diagonal magnetic moments\label{sec:GDMM}}

In this section, we study spin-flavor oscillations of neutrinos with great diagonal magnetic moments in a plane electromagnetic wave.

In some extensions of the standard model, e.g., in a minimally extended standard model, diagonal magnetic moments of neutrino are proportional to neutrino masses~\cite{DvoStu04}. The transition magnetic moment is much smaller than diagonal ones because of the Glashaw-Iliopoulos-Maiani (GIM) mechanism~\cite[pp.~461\textendash 479]{FukYan03}. Considering the situation of great diagonal magnetic moments, we can derive analytically the transition probability of neutrino spin-flavor oscillations in an electromagnetic wave for such neutrinos.

As in Sec.~\ref{sec:TMM}, we study the evolution of neutrino mass eigenstates. The Dirac equation for $\psi_a$ coincides with that in Eqs.~\eqref{eq:psiaeq} and~\eqref{eq:HaV}. However we should take that $V=0$ since we study the situation when $\mu_a \gg \mu$.

Since the transition magnetic moment is not taken into account, one has the following qualitative evolution of the neutrino spin and mass eigenstates. There will be independent spin precession within each mass eigenstate owing to the interaction of the neutrino magnetic moment $\mu_a$ with the electromagnetic field of the wave. The transition between different mass eigenstates is solely due to the presence of the nonzero vacuum mixing in Eq.~\eqref{eq:mattrans}.

Therefore, to find the transition probability for $\nu_{\beta\mathrm{L}}\to\nu_{\alpha\mathrm{R}}$ oscillations, we can use Eqs.~\eqref{eq:psisol}-\eqref{eq:vsol}, as well as the results of Ref.~\cite{DvoMaa07}, where spin-flavor oscillations of neutrinos with arbitrary magnetic moments were studied in the constant magnetic field. We assume that neutrinos are ultrarelativistic particles and adopt the quasiclassical approximation from the very beginning. Omitting some intermediate calculations, we just give the final result for the transition probability,
\begin{equation}\label{eq:Pgmua}
  P_{\nu_{\beta\mathrm{L}}\to\nu_{\alpha\mathrm{R}}}(t) =
  \sin^2(2\theta)
  \left[
    \frac{1}{4}(A_1 - A_2)^2 + A_1 A_2 \sin^2(\Phi_\mathrm{vac}t)
  \right],
\end{equation}
where
\begin{equation}\label{eq:A12}
  A_{1,2}(t) =
  \frac{\mu_{1,2} B_{0}}{\sqrt{\mu_{1,2}^{2}B_{0}^{2}+\omega^{2}/4}}
  \sin
  \left[
    \sqrt{\mu_{1,2}^{2}B_{0}^{2}+\omega^{2}/4}(1-\beta_{1,2})t
  \right],
\end{equation}
are the amplitudes of probabilities of spin oscillations within each mass eigenstate and $\Phi_\mathrm{vac} = \delta m^2/4p$ is the phase of flavor oscillations in vacuum.

If we set $\omega = 0$ and replace $(1-\beta_{1,2}) \to 1$ in Eq.~\eqref{eq:A12}, we get that $A_{1,2} = \sin(\mu_{1,2} B_0 t)$. Thus Eq.~\eqref{eq:Pgmua} reproduces the transition probability for $\nu_{\beta\mathrm{L}}\to\nu_{\alpha\mathrm{R}}$ oscillations in a constant transverse magnetic field derived in Ref.~\cite{DvoMaa07} for the case $\mu_a \gg \mu$.

Although the transition probability, obtained in Eqs.~\eqref{eq:Pgmua} and~\eqref{eq:A12} is  analytical, the situation when $\mu_a \gg \mu$ has very limited applicability from the point of view of phenomenology. The fact is that $\mu_a \sim 10^{-19}\mu_\mathrm{B}$ in the minimally extended standard model for reasonable values of neutrino masses~\cite{Ase11}. Thus $P_{\nu_{\beta\mathrm{L}}\to\nu_{\alpha\mathrm{R}}} \ll 1$ for any $t$. That is why we do not provide the correction to $P_{\nu_{\beta\mathrm{L}}\to\nu_{\alpha\mathrm{R}}} \ll 1$ owing to $\mu\neq 0$ since such a quantity is really negligible.

\section{Conclusion\label{sec:CONCL}}

In conclusion, we mention that, in the present work, we have studied
spin and spin-flavor oscillations of Dirac neutrinos in a plane electromagnetic
wave with the circular polarization. Our analysis was based on solving
the initial condition problem for the system of flavor neutrinos interacting
with external fields. This method was reviewed in Ref.~\cite{Dvo11}.
The neutrino interaction with the electromagnetic wave is owing to
the presence of the nonzero neutrino magnetic moments. This interaction
was accounted for by using the exact solution of the Dirac-Pauli equation
for massive fermions found in Ref.~\cite{BagGit90}.

In Sec.~\ref{sec:SPINOSC}, we have considered neutrino spin oscillations
in a plane electromagnetic wave within one neutrino mass eigenstate.
We have obtained the exact expression for the transition probability
for $\nu_{\mathrm{L}}\to\nu_{\mathrm{R}}$ oscillations in Eq.~(\ref{eq:PLR}).
This transition probability turns out to depend on both time and the
spatial coordinate. However, for an ultrarelativistic neutrino, which
is localized along its trajectory, the spatial coordinate dependence
is transformed to the dependence on time only. In this case, we have
reproduced the transition probability obtained in Ref.~\cite{EgoLobStu00},
where the quasiclassical approach for the description of the neutrino
spin evolution in electromagnetic fields was developed.

Then, in Secs.~\ref{sec:TMM}-\ref{sec:GDMM}, we have generalized
the approach in Sec.~\ref{sec:SPINOSC} to treat neutrino spin-flavor
oscillations. In Sec.~\ref{sec:TMM}, we have started with the
analysis of the situation when only a great transition magnetic moment
is present in the mass eigenstates basis. In this situation, one can
find the analytic expression for the transition probability; cf. Eq.~(\ref{eq:PbetaLalphaR}).
Comparing the transition probabilities in Fig.~\ref{fig:PTMM} with
the results of Ref.~\cite{Dvo08}, one can see that spin-flavor oscillations
of Dirac neutrinos in an electromagnetic wave are analogous to oscillations
in a twisting magnetic field.

The expression for the transition probability in Eq.~(\ref{eq:PbetaLalphaR})
turns out to differ from the analogous result obtained in Ref.~\cite{EgoLobStu00}; cf. Eq.~\eqref{eq:Pwrong}.
This discrepancy is due to the incorrect treatment of transitions
between different neutrino flavor eigenstates in Ref.~\cite{EgoLobStu00}.
As we have mentioned above, the neutrino spin evolution within one
neutrino mass eigenstate was accounted for correctly in Ref.~\cite{EgoLobStu00}.
Thus, in the present work, we have provided the valid generalization
of the results of Ref.~\cite{EgoLobStu00} for the description of
neutrino spin-flavor oscillations in an electromagnetic wave.

Then, in Sec.~\ref{sec:TMM}, we have discussed a possibility of an astrophysical
application of the obtained results. We have studied $\nu_{e\mathrm{L}}\to\nu_{\mu\mathrm{R}}$
oscillations in an electromagnetic wave emitted by a highly magnetized
NS. We found that the upper bound on the averaged transition probability can reach $35\%$; cf. Fig.~\ref{fig:PTMM}. If the amplitude of an electromagnetic wave is less than $10^{18}\,\text{G}$, which is likely to be the case in realistic astrophysical media, the maximal transition probability can be found in Eq.~\eqref{eq:PtrsmallB}.

In Sec.~\ref{sec:DMM}, we have briefly discussed the
influence of the diagonal magnetic moments on spin-flavor oscillations
in an electromagnetic wave. We have assumed that diagonal magnetic
moments, $\mu'\sim10^{-14}\mu_{\mathrm{B}}$, are much smaller than
the transition one, $\mu\sim10^{-11}\mu_{\mathrm{B}}$. This value
of $\mu'$ is in agreement with the upper bound on diagonal magnetic
moments of Dirac neutrinos revealed in Ref.~\cite{Bel05}. One can
see in Figs.~\ref{fig:PDMM12} and~\ref{fig:PDMM13} that, if diagonal magnetic moments
are taken into account, the averaged transition probability is less
than in the case when only a transition magnetic moment is accounted
for. Besides $\nu_{e\mathrm{L}}\to\nu_{\mu\mathrm{R}}$ oscillations channel, we have studied other types of neutrino oscillations like $\nu_{e\mathrm{L}}\to\nu_{\mu\mathrm{L}}$, $\nu_{e\mathrm{L}}\to\nu_{e\mathrm{R}}$, and $\nu_{e\mathrm{L}}\to\nu_{e\mathrm{L}}$ in Sec.~\ref{sec:DMM}; cf. Figs.~\ref{fig:eLmuL13}-\ref{fig:eLeL13}.

Finally, in Sec.~\ref{sec:GDMM}, we have provided the analytical transition probability for neutrino spin-flavor oscillations in an electromagnetic wave in case when diagonal magnetic moments dominate over transition one; cf. Eqs.~\eqref{eq:Pgmua} and~\eqref{eq:A12}. This situation can take place, e.g., in the minimally extended standard model, where the transition magnetic moment is suppressed by the GIM mechanism~\cite[pp.~461\textendash 479]{FukYan03}. However, this our result is of limited phenomenological applicability since the diagonal magnetic moments are extremely small in this model.

\section*{Acknowledgments}

%\begin{acknowledgments}
This work was partially supported by the Competitiveness Improvement
Program at the Tomsk State University, RFBR (Research Project No.
18-02-00149a), and the Foundation for the Advancement of Theoretical Physics and
Mathematics ``BASIS''. I am also thankful to S.~Dvornikov for the assistance
in numerical simulations.
%\end{acknowledgments}

\appendix

\section{Matrix elements of spin interaction\label{sec:MATREL}}

In this appendix, we compute the matrix elements of the neutrino spin
interaction with a plane circularly polarized electromagnetic wave
between two component spinors corresponding to different mass eigenstates.

First, we replace $\mu\to\mu_{a}$ in Eq.~(\ref{eq:vsol}). Then,
we define $v_{as}(t)$ corresponding to $v_{0s}$, where $v_{0+}^{\mathrm{T}}=(1,0)$
and $v_{0-}^{\mathrm{T}}=(0,1)$. Basing on the unitarity of the evolution
of the spinors in Eq.~(\ref{eq:vsol}), one gets that $v_{as}^{\dagger}(t)v_{as'}(t)=v_{0s}^{\dagger}v_{0s'}=\delta_{ss'}$,
i.e. the spinors $v_{as}(t)$ are orthogonal.

After a lengthy but straightforward calculation we obtain the matrix
elements $v_{as}^{\dagger}(t) (\bm{\sigma} \mathbf{B}) \\ \times v_{bs'}(t)$,
$a\neq b$, which enter to Eq.~(\ref{eq:aeqraw}), in the following
form:
\begin{align}\label{eq:me1}
  v_{1\pm}^{\dagger}(\bm{\sigma}\mathbf{B})v_{2\pm}= &
  \frac{B_{0}}
  {\sqrt{\mu_{1}^{2}B_{0}^{2}+\omega^{2}/4}\sqrt{\mu_{2}^{2}B_{0}^{2}+\omega^{2}/4}}
  \displaybreak[1]
  \\
  & \times
  \bigg\{
    \pm(\mu_{1}+\mu_{2})B_{0}\frac{\omega}{2}
    \sin
    \left[
      \sqrt{\mu_{1}^{2}B_{0}^{2}+\omega^{2}/4}\phi
    \right]
    \sin
    \left[
      \sqrt{\mu_{2}^{2}B_{0}^{2}+\omega^{2}/4}\phi
    \right]
    \displaybreak[1]
    \nonumber
    \\
    & +
    \mathrm{i}\mu_{2}B_{0}\sqrt{\mu_{1}^{2}B_{0}^{2}+\omega^{2}/4}
    \cos
    \left[
      \sqrt{\mu_{1}^{2}B_{0}^{2}+\omega^{2}/4}\phi
    \right]
    \sin
    \left[
      \sqrt{\mu_{2}^{2}B_{0}^{2}+\omega^{2}/4}\phi
    \right]
    \displaybreak[1]
    \nonumber
    \\
    & -\mathrm{i}\mu_{1}B_{0}\sqrt{\mu_{2}^{2}B^{2}+\omega^{2}/4}
    \sin
    \left[
      \sqrt{\mu_{1}^{2}B_{0}^{2}+\omega^{2}/4}\phi
    \right]
    \cos
    \left[
      \sqrt{\mu_{2}^{2}B_{0}^{2}+\omega^{2}/4}\phi
    \right]
  \bigg\} ,
  \nonumber
  \displaybreak[1]
  \\
  v_{1-}^{\dagger}(\bm{\sigma}\mathbf{B})v_{2+}= &
  \frac{B_{0}}
  {\sqrt{\mu_{1}^{2}B_{0}^{2}+\omega^{2}/4}\sqrt{\mu_{2}^{2}B_{0}^{2}+\omega^{2}/4}}  
  \label{eq:me2}
  \displaybreak[1]
  \\
   & \times
  \bigg\{
    \left(
      \mu_{1}\mu_{2}B_{0}^{2}-\frac{\omega^{2}}{4}
    \right)
    \sin
    \left[
      \sqrt{\mu_{1}^{2}B_{0}^{2}+\omega^{2}/4}\phi
    \right]
    \sin
    \left[
      \sqrt{\mu_{2}^{2}B_{0}^{2}+\omega^{2}/4}\phi
    \right]
    \nonumber
    \displaybreak[1]
    \\
    & +
    \sqrt{\mu_{1}^{2}B_{0}^{2}+\omega^{2}/4}\sqrt{\mu_{2}^{2}B_{0}^{2}+\omega^{2}/4}
    \notag
    \displaybreak[1]
    \\
    & \times
    \cos
    \left[
      \sqrt{\mu_{1}^{2}B_{0}^{2}+\omega^{2}/4}\phi
    \right]
    \cos
    \left[
      \sqrt{\mu_{2}^{2}B_{0}^{2}+\omega^{2}/4}\phi
    \right]
    \nonumber
    \displaybreak[1]
    \\
    & +
    \mathrm{i}\frac{\omega}{2}\sqrt{\mu_{1}^{2}B_{0}^{2}+\omega^{2}/4}
    \cos
    \left[
      \sqrt{\mu_{1}^{2}B_{0}^{2}+\omega^{2}/4}\phi
    \right]
    \sin
    \left[
      \sqrt{\mu_{2}^{2}B_{0}^{2}+\omega^{2}/4}\phi
    \right]
    \nonumber
    \displaybreak[1]
    \\
    & +
    \mathrm{i}\frac{\omega}{2}\sqrt{\mu_{2}^{2}B_{0}^{2}+\omega^{2}/4}
    \sin
    \left[
      \sqrt{\mu_{1}^{2}B_{0}^{2}+\omega^{2}/4}\phi
    \right]
    \sin
    \left[
      \sqrt{\mu_{2}^{2}B_{0}^{2}+\omega^{2}/4}\phi
    \right]
  \bigg\},
  \nonumber
  \displaybreak[1]
  \\
  v_{1+}^{\dagger}(\bm{\sigma}\mathbf{B})v_{2-}= &
  \frac{B_{0}}
  {\sqrt{\mu_{1}^{2}B_{0}^{2}+\omega^{2}/4}\sqrt{\mu_{2}^{2}B_{0}^{2}+\omega^{2}/4}}  
  \label{eq:me3}
  \displaybreak[1]
  \\
  & \times
  \bigg\{
    \left(
      \mu_{1}\mu_{2}B_{0}^{2}-\frac{\omega^{2}}{4}
    \right)
    \sin
    \left[
      \sqrt{\mu_{1}^{2}B_{0}^{2}+\omega^{2}/4}\phi
    \right]
    \sin
    \left[
      \sqrt{\mu_{2}^{2}B_{0}^{2}+\omega^{2}/4}\phi
    \right]
    \nonumber
    \\
    & +
    \sqrt{\mu_{1}^{2}B_{0}^{2}+\omega^{2}/4}\sqrt{\mu_{2}^{2}B_{0}^{2}+\omega^{2}/4}
    \notag
    \\
    & \times
    \cos
    \left[
      \sqrt{\mu_{1}^{2}B_{0}^{2}+\omega^{2}/4}\phi
    \right]
    \cos
    \left[
      \sqrt{\mu_{2}^{2}B_{0}^{2}+\omega^{2}/4}\phi
    \right]
    \nonumber
    \\
    & -
    \mathrm{i}\frac{\omega}{2}\sqrt{\mu_{1}^{2}B_{0}^{2}+\omega^{2}/4}
    \cos
    \left[
      \sqrt{\mu_{1}^{2}B_{0}^{2}+\omega^{2}/4}\phi
    \right]
    \sin
    \left[
      \sqrt{\mu_{2}^{2}B_{0}^{2}+\omega^{2}/4}\phi
    \right]
    \nonumber
    \\
   & -
    \mathrm{i}\frac{\omega}{2}\sqrt{\mu_{2}^{2}B_{0}^{2}+\omega^{2}/4}
    \sin
    \left[
      \sqrt{\mu_{1}^{2}B_{0}^{2}+\omega^{2}/4}\phi
    \right]
    \sin
    \left[
      \sqrt{\mu_{2}^{2}B_{0}^{2}+\omega^{2}/4}\phi
    \right]
  \bigg\},
  \nonumber
\end{align}
where $\phi=t-z$.

In Sec.~\ref{sec:DMM}, we are interested in the situation of
small diagonal magnetic moments: $\mu_{a}\ll\mu$. Expanding Eqs.~(\ref{eq:me1})-(\ref{eq:me3})
in $\mu_{a}B_{0}$ we get that only $v_{1\pm}^{\dagger}(\bm{\sigma}\mathbf{B})v_{2\pm}$
contains the contribution linear in $\mu_{a}$,
\begin{equation}\label{eq:veps}
  v_{1\pm}^{\dagger}(\bm{\sigma}\mathbf{B})v_{2\pm}=\frac{B_{0}^{2}}{\omega}
  \left[
    \pm
    \left(
      \mu_{1}+\mu_{2}
    \right)
    \mp\mu_{1}e^{\pm i\omega\phi}\mp\mu_{2}e^{\mp i\omega\phi}
  \right].
\end{equation}
The expressions for $v_{1\pm}^{\dagger}(\bm{\sigma}\mathbf{B})v_{2\mp}$
in Eqs.~(\ref{eq:me2}) and~(\ref{eq:me3}) are $\sim\mu_{a}^{2}$.
Hence we should use the zero order approximation in Eq.~(\ref{eq:meanv})
for them.

\end{document}